\documentclass[aps,prd,reprint,amsmath,amssymb,superscriptaddress,longbibliography,floatfix]{revtex4-2}
\usepackage[T1]{fontenc}
\usepackage{lmodern}
\usepackage[utf8]{inputenc}
\usepackage{bm}
\usepackage{braket}
\usepackage{amsthm}
\usepackage{tikz}
\usepackage{graphicx}
\usepackage{booktabs}
\usepackage{pgfplots}
\pgfplotsset{compat=1.18}
\usepackage[colorlinks=true,linkcolor=blue,citecolor=blue,urlcolor=blue]{hyperref}
\newcommand{\Tr}{\operatorname{Tr}}
\newcommand{\id}{\mathbf{1}}
\newcommand{\ii}{\mathrm{i}}
\newcommand{\dd}{\mathrm{d}}
\newcommand{\DU}{D_{\mathrm U}}
\newcommand{\gBKM}{g^{\mathrm{BKM}}}
\newcommand{\gSLD}{g^{\mathrm{SLD}}}
\newcommand{\gFS}{g^{\mathrm{FS}}}
\newtheorem{proposition}{Proposition}
\newtheorem{lemma}{Lemma}
\theoremstyle{definition}

\theoremstyle{remark}
\newtheorem{remark}{Remark}
\begin{document}
\title{A Multi-Affine Geometric Framework for Quantum Nonlocality: Unifying Berry Phases, Entanglement, and Coherence}
\author{Shoshauna Gauvin}
\email{s3gauvin@uwaterloo.ca}
\affiliation{Relativistic Quantum Information (RQI) Lab, Department of Physics, University of Waterloo, Waterloo, Ontario, Canada}
\begin{abstract}
We relate Berry-phase control and retained entropy geometry to recoverable coherence, entanglement, and Clauser-Horne-Shimony-Holt (CHSH) correlations. For a depolarized Bell pair subject to recorded phase noise, we determine exact CHSH optima when Bob retains his record and when he compresses it with his qubit into one qubit by any deterministic quantum channel. Some records preserve a violation that every such compression loses. The retained transverse metric gives sharp resource intervals. Nevertheless, for every finite derivative order, two finite readouts of the same noise can match all pulled-back monotone-metric and relative-entropy derivatives at faithful diagonal inputs while giving opposite recovered CHSH outcomes. Complete transverse entropy response determines the reliability law. A finite transitionless Berry loop realizes the channel; zero-expectation conditional control errors give a quadratic reduced-channel accuracy certificate. Transport, thermal, and accelerated-detector applications specialize established channel criteria under explicit preparation and accuracy assumptions. These model-dependent certificates supply neither a universal entanglement lifetime nor a detector-independent acceleration bound.
\end{abstract}
\maketitle

\section{Introduction}
\label{sec:intro}
Which correlations remain accessible after phase noise depends on the
preparation, control, and retained observations. Berry transport,
statistical distinguishability, and entanglement describe different
structures: projective geometry concerns rays and phase
\cite{BrodyHughston2001,Berry1984,Simon1983}, information geometry supplies
dual affine connections~\cite{AmariNagaoka2000}, and a subsystem split
defines bipartite correlations. We compare them on common preparations
with specified channels and measurements.

The controlled-completion framework~\cite{GauvinCompletion2026} organizes
statistical, tangent-bundle, and transport structures on a supplied
quantum carrier. Here multi-affine compares a specified dual-affine
structure with these induced geometries; unifying refers to their
operational relations, with quantum kinematics and subsystem structure
supplied as physical inputs. Discarded distinguishability is a contrast
loss, whose Hessian we compare with recoverable resources.

The phase-record model yields two exact results. For a depolarized Bell
pair, we optimize CHSH over every deterministic compression of Bob's
qubit and record into one qubit and over measurements retaining his
record. Their strict separation quantifies the output restriction,
providing a specified-state comparison related to measurement-relative
compression~\cite{Bluhm2018Compression}. Second, for any prescribed finite
order, two readouts of the same noise match all pulled-back monotone-metric
derivatives at faithful diagonal references and pulled-back contrast
derivatives at diagonal pairs, yet have opposite optimized CHSH outcomes after
compression. This combines established reliability moments
\cite{Reeves2024Moments,Goela2020Blackwell} with quantum metric pullbacks
and the recovery optimum. Sharp metric-only intervals and an explicit
curvature-matched example complement the general construction.

The ingredients have established origins: the probability-phase
realization adapts Molitor~\cite{Molitor2012}, the projector response is
the quantum geometric tensor~\cite{ProvostVallee1980}, and the
faithful-orbit factors specialize monotone kernels~\cite{PetzSudar1996}.
Geometric-phase control of Bell-type correlations
\cite{Sponar2010,JhaMalikBoyd2008}, transitionless gates
\cite{Berry2009Transitionless,QiJing2020}, environment-assisted phase
correction~\cite{GregorattiWerner2003,TrendelkampSchroer2011}, and
syndrome-compression information loss~\cite{Sheng2026Syndrome} are also
established. Our conditional-control certificate bounds reduced-channel
error quadratically when the error expectation vanishes along the
transported ray. The known isoholonomic bound
\cite{HornedalSonnerborn2023GeometricPhase,Sonnerborn2024Isoholonomic,Sonnerborn2025Tight} then quantifies how
a shorter realization improves this accuracy budget.

The applications use standard phase averaging~\cite{Cywinski2008} for a
finite-memory Gaussian model, and known thermal and accelerated-detector
criteria~\cite{KofmanKorotkov2008,Ali2009,Filippov2018,Tian2013,HuYu2015,LamiGiovannetti2015}.
They distinguish preparation-specific loss from channel entanglement
breaking and require output-error margins; no microscopic detector-error
theorem is supplied.

We assume complex Hilbert spaces, normalized positive states, the Born
rule, and specified local observable algebras. Each dynamical model
supplies its Hamiltonian, clock, and couplings. Relative entropy uses
natural logarithms; resource entropies use bits. State-space coordinates
are distinct from physical time.

Section~\ref{qm:sec:model} fixes the preparation and resource conventions;
Sec.~\ref{sec:phase-operations} gives the loop and control certificate;
Sec.~\ref{eg:sec_entropy} establishes the record and geometric results.
Sections~\ref{dyn:section}-\ref{uc:sec} treat transport, thermal dynamics,
accelerated detectors, and carrier scales. The appendices give geometric
conventions and proofs.

\section{Quantum model, geometry, and measured resources}
\label{qm:sec:model}

Fix the physical tensor product $\mathcal H=\mathbb C^2_A\otimes\mathbb C^2_B$, its computational basis, local observable algebras, and Born rule.

Consider the normalized family
\begin{equation}
 \ket{\psi(\vartheta,\phi)}
 =\cos\vartheta\ket{00}
   +e^{\ii\phi}\sin\vartheta\ket{11},
 \label{qm:eq:state}
\end{equation}
with $0<\vartheta<\pi/2$, $\phi$ modulo $2\pi$, $P=\ket\psi\bra\psi$,
$p=\sin^2\vartheta$, and $q=1-p$. Endpoint limits make $\phi$ redundant;
ordered Schmidt weights restrict $\vartheta$ to $[0,\pi/4]$.
Here $\vartheta$ is a control parameter and $t$ denotes physical time.

\subsection{Preparation and a gapped transport protocol}
\label{qm:subsec:preparation}

A local rotation followed by CNOT prepares the family:
\begin{align}
 &\operatorname{CNOT}\bigl[
  (\sqrt q\ket0+e^{\ii\phi}\sqrt p\ket1)_A\ket0_B
  \bigr]\nonumber\\
 &\hspace{28mm}=\sqrt q\ket{00}+e^{\ii\phi}\sqrt p\ket{11}.
 \label{qm:eq:cnot}
\end{align}
The joint CNOT can create entanglement from the coherent product input.

To realize geometric transport, supply the parent Hamiltonian
\begin{equation}
 H(\vartheta,\phi)=\Delta_E[\id-P(\vartheta,\phi)],
 \qquad \Delta_E>0.
 \label{qm:eq:parent}
\end{equation}
This supplied joint Hamiltonian has a unique zero-energy ground ray and
a three-dimensional excited subspace at gap $\Delta_E$. Slow cyclic
control gives adiabatic transport with zero ground-state dynamical phase;
a scalar energy shift requires interferometric phase calibration.

\subsection{Quantum metric and Berry curvature}
\label{qm:subsec:geometry}

For \(P=\ket{\psi}\bra{\psi}\), a normalized local section gives
\begin{equation}
 Q_{ij}
 =\bra{\partial_i\psi}(\id-P)\ket{\partial_j\psi},
 \qquad
 \gFS_{ij}=\operatorname{Re}Q_{ij}.
 \label{eg:ray_qgt}
\end{equation}
With the convention \(A_i=\ii\langle\psi|\partial_i\psi\rangle\),
\begin{equation}
 \mathcal F_{ij}
 =\partial_iA_j-\partial_jA_i
 =-2\operatorname{Im}Q_{ij}.
 \label{eg:ray_curvature}
\end{equation}
Berry phase is
the holonomy of this line connection \cite{Berry1984,Simon1983}.

For coordinate order $(\vartheta,\phi)$, these definitions give
\begin{align}
 \dd s_{FS}^2&=\dd\vartheta^2
  +\tfrac14\sin^2(2\vartheta)\,\dd\phi^2,\nonumber\\
 A&=-\sin^2\vartheta\,\dd\phi,\nonumber\\
 \mathcal F&=-\sin(2\vartheta)\,\dd\vartheta\wedge\dd\phi.
 \label{qm:eq:geometry}
\end{align}
For a once-wound, positively oriented $\phi$ loop at fixed $\vartheta$, the state vector in Eq.~\eqref{qm:eq:state} is single-valued and
\begin{equation}
 \gamma_B=-2\pi\sin^2\vartheta\pmod{2\pi}.
 \label{qm:eq:berry-loop}
\end{equation}
Equation~\eqref{qm:eq:parent} supplies its adiabatic realization~\cite{Berry1984}; orientation and winding change the accumulated phase while the instantaneous entanglement remains fixed.

\subsection{The phase generator and an observable geometric phase}
\label{qm:subsec:phase-readout}

The phase coordinate has a specified physical generator,
\begin{equation}
 G=\id_A\otimes\ket1_B\bra1,
 \qquad \partial_\phi\ket\psi=\ii G\ket\psi.
 \label{qm:eq:phase-generator}
\end{equation}
Thus $A_\phi=-\langle G\rangle=-p$ and
$\gFS_{\phi\phi}=\operatorname{Var}(G)=pq$, with both the generator and
metric component tied to the parameter normalization.

Whole-state cyclic transport changes only an overall phase, measurable
by coherent comparison [Eq.~\eqref{op:fringe}]. The conditional loop in
Sec.~\ref{op:finite-loop} implements a local relative-phase gate,
preserving entanglement and optimized CHSH. General local unitaries
can change fixed-basis coherence.

\begin{figure}[t]
\centering
\begin{tikzpicture}
\begin{axis}[width=\linewidth,height=5.5cm,xmin=0,xmax=.25,ymin=0,ymax=1.04,
 xlabel={$\vartheta/\pi$},ylabel={Resource or normalized phase},
 xtick={0,.125,.25},xticklabels={$0$,$1/8$,$1/4$},
 tick label style={font=\small},label style={font=\small},
 legend columns=-1,legend style={font=\footnotesize,at={(.5,1.03)},anchor=south,draw=none},
 grid=major,grid style={gray!15}]
\addplot[blue!75!black,thick] table[x=theta_pi,y=concurrence]{figures/pure_resources.dat};
\addlegendentry{$\mathcal C$}
\addplot[red!75!black,thick,dashed] table[x=theta_pi,y=entropy]{figures/pure_resources.dat};
\addlegendentry{$E=C_r(P)$ (bits)}
\addplot[teal!70!black,thick,dashdotted] table[x=theta_pi,y=phase]{figures/pure_resources.dat};
\addlegendentry{$|\gamma_B|/\pi$}
\end{axis}
\end{tikzpicture}
\caption{Schmidt-family resources with ordered weights. The loop winds $\phi$ once positively; the phase representative is $\gamma_B\in[-\pi,0]$. Entanglement entropy equals global relative-entropy coherence in the computational basis; both local coherences vanish.}
\label{fig:pure}
\end{figure}

\subsection{Entanglement and coherence}
\label{qm:subsec:resources}

The reduced states are
\begin{equation}
 \rho_A=\rho_B=\operatorname{diag}(q,p).
 \label{qm:eq:marginals}
\end{equation}
Using base-2 logarithms for resource entropies, the entanglement entropy and pure-state concurrence are
\begin{equation}
 E=h_2(p),\qquad \mathcal C=2\sqrt{pq}=\sin(2\vartheta),
 \label{qm:eq:resources}
\end{equation}
where $h_2(p)=-p\log_2p-(1-p)\log_2(1-p)$.

For a fixed reference basis with dephasing $\Delta$, define coherence in
bits by~\cite{Baumgratz2014}
\begin{equation}
 C_r(\rho)=S_2(\Delta\rho)-S_2(\rho),
 \label{ch:coherence}
\end{equation}
where $S_2(\rho)=-\Tr\rho\log_2\rho$. We use the computational basis
for the present preparation.

For computational-basis coherence, Eqs.~\eqref{ch:coherence}
and~\eqref{qm:eq:marginals} give
\begin{equation}
 C_r(\rho_A)=C_r(\rho_B)=0,
 \qquad C_r(P)=h_2(p)=E.
 \label{qm:eq:coherence-values}
\end{equation}
This equality is specific to the family and computational basis.

CNOT converts the input's local coherence $h_2(p)$ into entanglement
$h_2(p)$, leaving zero local coherence and preserving total computational-
basis coherence. This joint basis permutation is incoherent but nonlocal,
realizing the established conversion~\cite{Streltsov2015}.

\subsection{Bell correlations and readout conventions}
\label{qm:sec:bell}
\label{dyn:conventions}

Let $A_x$ and $B_y$ be local Hermitian observables with $A_x^2=B_y^2=\id$ and $x,y\in\{0,1\}$. Define
\begin{align}
 E_{xy}&=\Tr[\rho(A_x\otimes B_y)],\nonumber\\
 S&=E_{00}+E_{01}+E_{10}-E_{11}.
 \label{qm:eq:chsh}
\end{align}
A setting-independent hidden-variable model with local factorization
obeys $|S|\leq2$~\cite{Bell1964,CHSH1969}. For a normalized positive
quantum state and the local observables above, the standard operator
bound is $|S|\leq2\sqrt2$~\cite{Tsirelson1980}.

For a two-qubit state let $T_{jk}=\Tr(\rho\sigma_j\otimes\sigma_k)$,
and write its singular values as $s_1\ge s_2\ge s_3$. The optimum over
nondegenerate local projective spin observables is \cite{Horodecki1995}
\begin{equation}
 S_{\mathrm{spin}}(\rho)=2\sqrt{s_1^2+s_2^2}.
 \label{dyn:spin_optimum}
\end{equation}
Allowing all binary qubit measurements gives
$S_{\max}=\max\{2,S_{\mathrm{spin}}\}$. Indeed, extreme binary qubit observables are Pauli
directions or $\pm\id$; a trivial setting cannot violate CHSH.
Thus $S_{\mathrm{spin}}>2$ is the criterion used below; its failure
establishes neither separability nor locality for every Bell protocol.
Unless stated otherwise, each storage time uses a fresh preparation
with single-copy, unconditioned measurements and reoptimized settings.
Losses and conditional experiments follow the instrument conventions in
Sec.~\ref{tr:detectors}.

For Eq.~\eqref{qm:eq:state} at $\phi=0$, this correlation matrix is
\begin{equation}
 T=\operatorname{diag}(\mathcal C,-\mathcal C,1).
 \label{qm:eq:corr-matrix}
\end{equation}
Choose
\begin{align}
 A_0&=\sigma_z,\qquad A_1=\sigma_x,\nonumber\\
 B_0&=\cos\alpha\,\sigma_z+\sin\alpha\,\sigma_x,\nonumber\\
 B_1&=\cos\alpha\,\sigma_z-\sin\alpha\,\sigma_x.
 \label{qm:eq:optimal-settings}
\end{align}
These settings give $S=2(\cos\alpha+\mathcal C\sin\alpha)$; choosing
$\tan\alpha=\mathcal C$ attains the Horodecki optimum
\cite{Horodecki1995}:
\begin{equation}
 S_{\max}=2\sqrt{1+\mathcal C^2}.
 \label{qm:eq:optimal-chsh}
\end{equation}
At maximal entanglement, $\alpha=\pi/4$ saturates the bound.

For $V_\phi=\operatorname{diag}(1,e^{\ii\phi})$, rotating Bob's observables to $V_\phi B_yV_\phi^\dagger$ compensates the preparation phase and preserves the optimum. Keeping Eq.~\eqref{qm:eq:optimal-settings} fixed instead gives
\begin{equation}
 S(\phi;\alpha)=2[\cos\alpha+\mathcal C\cos\phi\sin\alpha].
 \label{qm:eq:fixed-phase}
\end{equation}
Thus the relative phase changes fixed-setting correlations while
preserving entanglement and optimized CHSH. Experimental precedents
include geometric-phase compensation in neutron spin-path
contextuality tests~\cite{Sponar2010} and control of photon-pair
energy-time Bell correlations~\cite{JhaMalikBoyd2008}.

Within this specific preparation family,
\begin{equation}
 \mathcal C^2=4\gFS_{\phi\phi},\qquad
 S_{\max}=2\sqrt{1+4\gFS_{\phi\phi}}.
 \label{qm:eq:model-relations}
\end{equation}
These comparisons require the specified phase generator, parameter normalization, and physical tensor product; Sec.~\ref{sec:limits} gives the intrinsic counterexample.

Figure~\ref{fig:pure} compares the resources with the Berry phase for the specified loop.

\subsection{Why geometry alone does not fix nonlocality}
\label{sec:limits}
The single-qubit ray
\begin{equation}
 \ket{\chi(\vartheta,\phi)}
 =\cos\vartheta\ket0+e^{\ii\phi}\sin\vartheta\ket1
 \label{lim:single-qubit}
\end{equation}
has the same intrinsic quantum geometric tensor and Berry connection
as its isometric image $\cos\vartheta\ket{00}+e^{\ii\phi}\sin\vartheta\ket{11}$.
Only the target carries the bipartition needed for Schmidt rank and Bell
correlations; intrinsic ray geometry cannot reconstruct it. Local
unitaries, entangling gates, and changes of subsystem factorization are
distinct operations.

Bell-local factorization with setting-independent weights obeys CHSH
regardless of hidden-variable geometry~\cite{Bell1964}. Fine's equivalent
criterion in this scenario is a joint distribution for all potential
outcomes~\cite{Fine1982}, not merely a record of actual outcomes and settings.

Positivity and no signalling also permit the noisy Popescu-Rohrlich correlations~\cite{PopescuRohrlich1994}
\begin{equation}
 p(a,b|x,y)=\frac{1+ab\,u(-1)^{xy}}4,
 \label{lim:pr}
\end{equation}
where $a,b\in\{\pm1\}$, $x,y\in\{0,1\}$, and $0<u<1$. These strictly positive tables have uniform marginals but $S=4u$, exceeding $2\sqrt2$ for $u>1/\sqrt2$. Regular statistical geometry and no signalling therefore leave quantum realizability undetermined.

\subsection{Resource certificates from state accuracy}
\label{op:certification}

For the control and detector applications, let $\rho$ be the model
two-qubit output and $\widetilde\rho$ the actual output, with independently
justified full trace-norm error
\begin{equation}
 \|\widetilde\rho-\rho\|_1\le\epsilon.
 \label{ur:eq:state-error}
\end{equation}
Thus $\epsilon$ bounds the norm, not half the norm. Put $\mu(\rho)=\lambda_{\min}(\rho^{T_B})$. Partial transposition preserves Hilbert-Schmidt norm, so eigenvalue perturbation gives
\begin{align}
 |\mu(\widetilde\rho)-\mu(\rho)|
 &\le\|(\widetilde\rho-\rho)^{T_B}\|_\infty\nonumber\\
 &\le\|\widetilde\rho-\rho\|_2\le\epsilon.
 \label{ur:eq:ppt-error}
\end{align}
Thus $\mu(\rho)<-\epsilon$ certifies entanglement and $\mu(\rho)>\epsilon$ certifies separability of $\widetilde\rho$ by the two-qubit PPT criterion~\cite{Peres1996,Horodecki1996}; the intervening band is undecided.

Every CHSH operator has norm at most $2\sqrt2$, whence optimization over the same observable class gives
\begin{equation}
 |S_{\mathrm{spin}}(\widetilde\rho)-S_{\mathrm{spin}}(\rho)|
 \le2\sqrt2\,\epsilon.
 \label{ur:eq:chsh-error}
\end{equation}
A model spin value above $2+2\sqrt2\epsilon$ therefore certifies violation; one below $2-2\sqrt2\epsilon$ excludes violation for the binary qubit CHSH scenario. The same bound holds for $S_{\max}=\max(2,S_{\mathrm{spin}})$, whose floor prevents the strict nonviolation certificate.

For a qubit-to-qubit channel $\mathcal E_\tau$, define its normalized Choi state using $\ket{\Phi^+}=(\ket{00}+\ket{11})/\sqrt2$:
\begin{equation}
 J(\mathcal E_\tau)=
 (\mathcal E_\tau\otimes\mathrm{id})(\ket{\Phi^+}\bra{\Phi^+}),
 \label{ur:eq:choi}
\end{equation}
The channel is entanglement breaking exactly when this state is separable~\cite[Theorem~4]{HorodeckiShorRuskai2003}. Thus the two-qubit PPT margins also certify the actual channel's entanglement-breaking status, provided Eq.~\eqref{ur:eq:state-error} bounds its Choi-state error. A diamond-norm channel bound suffices; a bound for one arbitrary input alone does not.

\begin{figure}[t]
\centering
\begin{tikzpicture}[x=1cm,y=1cm,
  every node/.style={font=\small},
  gate/.style={draw,fill=blue!4,align=center,inner sep=4pt}]
\node at (0,2.6) {$A$};
\draw[->] (.3,2.6) -- (7.25,2.6);
\node[anchor=west] at (7.25,2.6) {$A_x$};

\node at (0,1.8) {$B$};
\draw (.3,1.8) -- (1.8,1.8);
\draw (4,1.8) -- (5,1.8);
\draw[->] (5.9,1.8) -- (7.25,1.8);
\node[anchor=west] at (7.25,1.8) {$B_y$};

\node at (0,.9) {$|0\rangle_R$};
\draw (.4,.9) -- (1.8,.9);
\draw[->] (4,.9) -- (4.8,.9);
\node[anchor=west] at (4.8,.9) {discard $R$};

\draw[fill=blue!4] (1.8,.55) rectangle (4,2.15);
\node[align=center] at (2.9,1.35) {$H_{BR,j}(t)$\\ closed loop};
\node[gate] at (5.45,1.8) {$F_Y$};

\node at (0,-.05) {$J$};
\draw[->,dashed] (.25,.05) -- (1.2,.05)
  -- (1.2,1.3) -- (1.8,1.3);
\draw[->] (.3,-.05) -- (2.1,-.05);
\node[gate] at (3.15,-.05) {$p(Y|J)$};
\draw[->] (4.2,-.05) -- (5.45,-.05) -- (5.45,1.48);
\node[anchor=west] at (5.55,.35) {$Y$};
\end{tikzpicture}
\caption{Local control and readout. The classical orientation $J$ selects
Bob's loop; the ancillary qubit $R$ is discarded. An available record
$Y$ permits phase feedback $F_Y$ before the independently chosen Bell
settings. Omitting feedback gives the uncorrected phase channel.
All trials are retained. Section~\ref{rec:sec} treats imperfect records.}
\label{fig:phase-protocol}
\end{figure}

\section{Phase transport, coherent comparison, and discarded information}
\label{sec:phase-operations}
\subsection{Interferometric readout retains relative phase}
Let a control qubit label two coherent branches. Starting from $\ket+\bra+\otimes\rho$, apply
\begin{equation}
 W=\ket0\bra0\otimes U_0+\ket1\bra1\otimes U_1.
 \label{op:controlled}
\end{equation}
The control's off-diagonal matrix element is one half of
\begin{equation}
 \nu=\Tr(U_0\rho U_1^\dagger).
 \label{op:visibility}
\end{equation}
Applying the control phase shifter $\operatorname{diag}(e^{\ii\chi},1)$ before an $X$ measurement gives
\begin{equation}
 p_+(\chi)=\frac{1+\operatorname{Re}(e^{\ii\chi}\nu)}2.
 \label{op:fringe}
\end{equation}
Scanning $\chi$ measures $|\nu|$ and, when nonzero, its phase.
Matching endpoint rays and calibrated dynamical phases make this a
geometric-phase difference for eigenlines; for higher rank, it is a
state-dependent matrix trace. The coherent control retains scalar
phases: $U_j=e^{\ii\gamma_j}\id$ gives the identity conjugation channel
but $\nu=e^{\ii(\gamma_0-\gamma_1)}$. Discarding a classical branch
record alone supplies no coherent fringe. Gauge changes require the matching
endpoint and preparation transformations, leaving Eq.~\eqref{op:fringe}
invariant.

\subsection{Random phase and accessible resources}
A uniformly chosen local phase rotation, with its classical label
discarded, defines
\begin{equation}
 U_\pm=e^{\pm\ii\delta\sigma_z/2},\qquad
 \Lambda=\tfrac12(\operatorname{Ad}_{U_+}
                         +\operatorname{Ad}_{U_-}).
 \label{op:phase-channel}
\end{equation}
Here $\operatorname{Ad}_U(X)=UXU^\dagger$. Off-diagonal qubit entries
are multiplied by $\cos\delta$, while diagonal entries are unchanged.

For a two-qubit Schmidt preparation, local phase randomization on one subsystem multiplies the $\ket{00}\bra{11}$ coherence by $\kappa=\cos\delta$. The resulting state, in the ordered $\ket{00},\ket{11}$ subspace, is
\begin{equation}
 \rho_\kappa=
 \begin{pmatrix}
 c^2&\kappa cs\,e^{-\ii\phi}\\
 \kappa cs\,e^{\ii\phi}&s^2
 \end{pmatrix},\qquad
 c=\cos\vartheta,\quad s=\sin\vartheta.
 \label{op:dephased-schmidt}
\end{equation}
Its other two eigenvalues in the two-qubit space vanish. The eigenvalues of the displayed block are
\begin{equation}
 \lambda_\pm=\frac{1\pm\sqrt{1-4c^2s^2(1-\kappa^2)}}2.
 \label{op:dephased-eigenvalues}
\end{equation}
Consequently its computational-basis coherence is
\begin{equation}
 C_r(\rho_\kappa)=h_2(s^2)-h_2(\lambda_+).
 \label{op:dephased-coherence}
\end{equation}
The partial transpose has eigenvalues $c^2,s^2,\pm|\kappa|cs$. For $0<\vartheta<\pi/2$, the state is entangled exactly when $\kappa\ne0$~\cite{Peres1996,Horodecki1996}. Its optimal spin CHSH value is
\begin{equation}
 S_{\mathrm{spin}}(\rho_\kappa)
 =2\sqrt{1+\kappa^2\sin^2(2\vartheta)}.
 \label{op:dephased-chsh}
\end{equation}
The correlation singular values $(1,z,z)$, with $z=|\kappa|\sin2\vartheta$, give the optimum~\cite{Horodecki1995}. Every entangled member of this family violates optimized CHSH. Section~\ref{qm:subsec:noise} contrasts this with depolarization; the present formulas are density-matrix resource identities, including at the boundary.

\subsection{A finite controlled-loop realization}
\label{op:finite-loop}
Fluctuating-loop dephasing has theoretical and experimental
precedents~\cite{DeChiaraPalma2003,Berger2015}. Following geometric gates
with counterdiabatic control and cubic schedules~\cite{QiJing2020},
we use transitionless transport~\cite{Berry2009Transitionless} to specify
Bob's ancillary loop, its orientation record, and its output-error budget
(Fig.~\ref{fig:phase-protocol}).

\emph{Control protocol.}
Bob attaches a local ancillary qubit $R$, initially $\ket0_R$, to his
logical qubit $B$. For orientation $j=\pm1$, write
$\ket{n_j}=\cos(\Theta/2)\ket0_R+
 e^{\ii\varphi_j}\sin(\Theta/2)\ket1_R$ and $P_j=\ket{n_j}\bra{n_j}$.
Supply the conditional Hamiltonian
\begin{align}
 H_{BR,j}(t)&=\ket1_B\bra1\otimes h_j(t),\nonumber\\
 h_j(t)&=\Delta_R(\id_R-P_j)+h_{{\rm cd},j}(t),\nonumber\\
 h_{{\rm cd},j}(t)&=\ii\hbar[\dot P_j,P_j],\qquad\Delta_R>0.
 \label{op:loop-hamiltonian}
\end{align}
Every term is controlled coherently and conditionally, including the scalar part of
$\Delta_R(\id_R-P_j)$; omitting it introduces a relative dynamical phase
requiring calibration. The parent gap is $\Delta_R$, $h_{{\rm cd},j}$ is
the counterdiabatic control, and the logical $\ket0_B$ sector is stationary.

Choose $0<\beta<\pi$, $\tau_L>0$, $T_L=3\tau_L$, and
$f(s)=3s^2-2s^3$ for $0\le s\le1$. The $C^1$, piecewise-smooth path is
\begin{equation}
 (\Theta,\varphi_j)=
 \begin{cases}
 (\beta f(t/\tau_L),0),&0\le t\le\tau_L,\\
 (\beta,2\pi j f(t/\tau_L-1)),&\tau_L\le t\le2\tau_L,\\
 (\beta[1-f(t/\tau_L-2)],2\pi j),&2\tau_L\le t\le T_L.
 \end{cases}
 \label{op:loop-path}
\end{equation}
First derivatives vanish at each junction, so the counterdiabatic term
vanishes there. At the endpoints the parent term also annihilates the
prepared ancilla.

\emph{Logical action.}
Since $[\dot P_j,P_j]\ket{n_j}=
\ket{\dot n_j}-\ket{n_j}\langle n_j|\dot n_j\rangle$, Schrödinger's
equation transports the logical $\ket1_B$ branch exactly along $P_j$.
Both Hamiltonian terms have zero expectation in that branch. At $T_L$
the ancilla returns to $\ket0_R$ and the logical gate is
\begin{equation}
 V_j=\operatorname{diag}(1,e^{-\ii j\delta}),\qquad
 \delta=\pi(1-\cos\beta).
 \label{op:loop-gate}
\end{equation}
Indeed $\gamma_j=-\oint\sin^2(\Theta/2)\,\dd\varphi_j=-j\delta$;
the meridians contribute zero. The added control realizes the parent
eigenline's Berry holonomy exactly in finite time.
The constant local Hamiltonian
$H_j^{\rm dyn}=\hbar j\delta\ket1_B\bra1/T_L$ produces the same $V_j$.
Both implementations therefore give identical averaged channels and
resource curves. Geometric attribution rests on the controlled path and
its zero dynamical phase, not on these resource curves.

\begin{proposition}[Retained or discarded loop orientation]
\label{op:prop:loop-channel}
Applied to Eq.~\eqref{qm:eq:state}, the local gate changes the relative
phase to $\phi-j\delta$, preserving entanglement and optimized CHSH.
Choose both orientations with probability $1/2$. A retained record $J$ gives
$\frac12\sum_j\ket j\bra j_J\otimes
(\id_A\otimes V_j)P(\id_A\otimes V_j^\dagger)$.
Bob can undo $V_j$ using that record on every trial; entanglement across
$A|(B,R,J)$ is preserved. Discarding $J$ and the returned ancilla gives
exactly Eq.~\eqref{op:dephased-schmidt} with $\kappa=\cos\delta$.
This realizes Eq.~\eqref{op:phase-channel} up to branch scalars.
For $N$ independent choices, read out at $t_N=NT_L$:
\begin{align}
 \kappa_N&=(\cos\delta)^N,\nonumber\\
 \mathcal C_N&=|\kappa_N|\sin(2\vartheta),\nonumber\\
 S_{{\rm spin},N}&=2\sqrt{1+\kappa_N^2\sin^2(2\vartheta)}.
 \label{op:loop-resources}
\end{align}
\end{proposition}
Coherence follows from Eq.~\eqref{op:dephased-coherence} with
$\kappa_N$. At $\beta=\pi/3$ one discarded orientation erases the
Schmidt coherence and entanglement; at $\beta=\pi/2$ the two logical
gates agree and there is no loss. For $0<|\cos\delta|<1$, the envelope
has the stroboscopic decay constant
$\gamma_L=-\ln|\cos\delta|/T_L$, fixed by the loop distribution and
repetition time. This stroboscopic envelope supplies no continuous
Markov description during a loop.

Recovery from a classical record relates to entanglement hidden by
unavailable classical information~\cite{DArrigo2014Recovery}; that
ensemble-dependent quantity differs from the retained entropy metric.

\emph{Control cost.}
With $\|\cdot\|$ denoting operator norm, the ancillary ray metric gives
$\|h_{{\rm cd},j}\|=\hbar\sqrt{g_{FS}(\dot P_j,\dot P_j)}$. Thus
\begin{align}
 \hbar^{-1}\int_0^{T_L}\|h_{{\rm cd},j}\|\,\dd t
   &=\beta+\pi\sin\beta=:L_\beta,\nonumber\\
 \max_t\|h_{{\rm cd},j}\|
   &=\frac{3\hbar}{4\tau_L}\max\{\beta,2\pi\sin\beta\}.
 \label{op:loop-cost}
\end{align}
Shorter cycles require stronger fields.

\emph{Error certificate.}
If the implemented Hamiltonian differs by $E_j(t)$, the unitary Duhamel
identity and contraction under discarding $R,J$ give
\begin{align}
 \eta_H&=\max_j\hbar^{-1}\int_0^{T_L}\|E_j(t)\|\,\dd t,\nonumber\\
 \|\rho^{\rm act}_{AB,N}-\rho^{\rm id}_{AB,N}\|_1
   &\le\varepsilon_N:=\min\{2,2N\eta_H\}.
 \label{op:loop-error}
\end{align}
Assume ideal preparation, branch probabilities, and readout, with the
bound valid for every realized branch in every loop. A fractional
counterdiabatic error $E_j=\zeta\ket1_B\bra1\otimes h_{{\rm cd},j}$
gives $\eta_H=|\zeta|L_\beta$. Other errors need separate bounds;
no microscopic reservoir estimate follows.

For a Bell input, the ideal minimum partial-transpose eigenvalue is
$-|\kappa_N|/2$. Consequently $|\kappa_N|/2>\varepsilon_N$ certifies
entanglement despite the bounded error, and
$S_{{\rm spin},N}>2+2\sqrt2\,\varepsilon_N$ certifies CHSH violation.
These are the margins of Eqs.~\eqref{ur:eq:ppt-error} and~\eqref{ur:eq:chsh-error} applied to Eq.~\eqref{op:loop-error}.
For example, $\beta=\pi/4$, $N=1$, and $|\zeta|\le10^{-3}$ give
$\varepsilon_1\le0.00602$ and a guaranteed spin CHSH value above $2.321$;
$\tau_L=20\hbar/\Delta_R$ requires a peak counterdiabatic norm below
$0.167\Delta_R$. Direct integration of Eq.~\eqref{op:loop-hamiltonian}
with this amplitude error verifies the benchmark in the accompanying
code. At exact erasure the ideal partial transpose is on the separable
boundary, so a nonzero generic error bound cannot certify separability.

Calibrated local $SU(2)$ rotations and computational readout implement
the optimized settings. Calibration or use of the branch record is
fixed before choosing the Bell settings. All trials are retained;
ancillary leakage is traced out.

\subsection{Control action and a sharper implementation certificate}
\label{op:control-refinement}
The same initial ancilla and branch gates admit a shorter ray loop.
Let $\mathbf z=(0,0,1)$, $\mathbf m_\beta=(\sin\beta,0,\cos\beta)$,
and $\chi_j(t)=2\pi j f(t/T_L)$, with the preceding cubic $f$. Replace
Eq.~\eqref{op:loop-path} by
\begin{align}
 \mathbf n_j(t)&=\cos\beta\,\mathbf m_\beta
 +(\mathbf z-\cos\beta\,\mathbf m_\beta)\cos\chi_j(t)\nonumber\\
 &\quad +(\mathbf m_\beta\mathbin\times\mathbf z)\sin\chi_j(t),
 \qquad P_j=\tfrac12(\id_R+\mathbf n_j\cdot\boldsymbol\sigma).
 \label{op:circle-path}
\end{align}
This circle starts and ends at $\mathbf z$ with zero speed. Its solid
angle is $2j\delta$, so Eq.~\eqref{op:loop-hamiltonian} implements the
same $V_j$, $\kappa_N$, and ideal resource responses. Its action and
peak counterdiabatic norm are
\begin{align}
 L_*&=\pi\sin\beta=\sqrt{\delta(2\pi-\delta)},\nonumber\\
 K_*&=\frac{3\hbar\pi\sin\beta}{2T_L}.
 \label{op:circle-cost}
\end{align}
Since $\|h_{\rm cd}\|/\hbar=|\dot{\mathbf n}|/2$, the circle saturates
the established pure-state isoholonomic inequality
\cite{HornedalSonnerborn2023GeometricPhase,Sonnerborn2024Isoholonomic,Sonnerborn2025Tight}. Thus $L_*$ minimizes
the integrated counterdiabatic norm among closed ancillary rays with this
holonomy and transitionless control. Total Hamiltonian cost, hardware
restrictions, and arbitrary dynamical implementations require separate
optimizations.
At $\beta=\pi/4$ and $T_L=60\hbar/\Delta_R$, the action drops by
$26.1\%$ and the peak norm becomes $0.05554\Delta_R$, one third of
the preceding value. Equation~\eqref{op:loop-error} then gives
$\varepsilon_1\le0.004443$ and $S_{\rm spin}^{\rm act}>2.3256$
for $|\zeta|\le10^{-3}$.

\begin{proposition}[Reduced-channel error with zero ray expectation]
\label{op:prop:quadratic}
Keep the conditional form of Eq.~\eqref{op:loop-hamiltonian}, with the
logical $\ket0_B$ branch unchanged. Let the ancillary Hamiltonian
error $e_j(t)$ be Hermitian and satisfy
$\langle n_j(t)|e_j(t)|n_j(t)\rangle=0$ throughout every realized loop.
Assume the initial ancillary state is $\ket0_R\bra0$, with
$P_j(0)=P_j(T_L)=\ket0_R\bra0$ for every loop, together with ideal
preparation and readout, fixed orientation probabilities, and complete
ancillary discarding at the final endpoint. If
\[
 \hbar^{-1}\int_{\text{one loop}}\|e_j(t)\|\,\dd t\le\eta_H
\]
for every realized branch, then the ideal and actual reduced channels
on $B$ obey, for $N$ loops without an ancilla reset,
\begin{equation}
 \|\mathcal E_N^{\rm act}-\mathcal E_N^{\rm id}\|_\diamond
 \le \varepsilon_N^{\perp}:=\min\{2,(N\eta_H)^2/2\}.
 \label{op:cd-quadratic-error}
\end{equation}
The bound also holds with a retained classical record. With an ideal
ancilla reset to $\ket0_R\bra0$ after every loop, $N\eta_H^2/2$
replaces $(N\eta_H)^2/2$.
\end{proposition}
The diamond norm controls full trace-norm error for every input, including
an entangled reference. The proof in Appendix~\ref{op:quadratic-proof}
uses zero ray expectation and ancillary discarding to improve the
generic Duhamel estimate. B\'eny's weak-interaction Kraus expansion
shows that the initial environment expectation fixes the first-order
coherent term~\cite[Sec.~II, Eq.~(6)]{Beny2011Perturbative}. The refinement
here is the explicit finite, time-dependent diamond bound and its reset
distinction for this conditional channel.

For real fractional counterdiabatic error
$e_j(t)=\zeta_j(t)h_{{\rm cd},j}(t)$,
$|\zeta_j(t)|\le\zeta_{\max}$, the hypothesis holds and
$\eta_H=\zeta_{\max}L$, where $L$ is the chosen loop action.
The circular benchmark therefore gives
$\varepsilon_1^{\perp}\le2.468\times10^{-6}$ and
$S_{\rm spin}^{\rm act}>2.33825$.
Additional diagonal phase errors, logical leakage, preparation
errors, or imperfect readout require separate bounds; an arbitrary
Hamiltonian perturbation still uses Eq.~\eqref{op:loop-error}.
This is a certificate for a specified control error, rather than a
universal robustness property of geometric gates.

\section{Entropy geometry and retained distinguishability}
\label{eg:sec_entropy}
Fix a complex Hilbert space \(\mathcal H\simeq\mathbb C^n\) with its
Hermitian product and operator adjoint. Its faithful state manifold is
\begin{equation}
 \mathcal S_n^+
 =\{\rho=\rho^\dagger>0:\Tr\rho=1\},
 \qquad T_\rho\mathcal S_n^+=\mathrm{Herm}_0(\mathcal H).
 \label{eg:state_manifold}
\end{equation}
The entropy contrast determines a statistical geometry; a ray or spectral
projector determines a phase geometry. Hamiltonians, instruments, and
subsystem decompositions supply their physical interpretation
\cite{GauvinCompletion2026}.

\subsection{Umegaki contrast and the faithful quantum model}
\label{eg:subsec_umegaki}

On Eq.~\eqref{eg:state_manifold}, use negative von Neumann entropy and
Umegaki relative entropy, with natural logarithms:
\begin{align}
 F(\rho)&=\Tr(\rho\log\rho),\\
 \DU(\rho\Vert\sigma)
 &=\Tr\!\left[\rho(\log\rho-\log\sigma)\right].
 \label{eg:umegaki}
\end{align}
Trace-one tangent vectors eliminate the identity term in the
differential of \(F\). On trace-zero Hermitian tangents \(U,V\), its
Hessian is the Bogoliubov-Kubo-Mori metric,
\begin{align}
 \mathcal K_\rho(A)
 &=\int_0^1\rho^sA\rho^{1-s}\,\dd s,\\
 \gBKM_\rho(U,V)
 &=\Tr\!\left[U\mathcal K_\rho^{-1}(V)\right].
 \label{eg:bkm_kernel}
\end{align}
The inverse is well defined at faithful \(\rho\).
In a diagonal basis the inverse-kernel (metric) coefficient is
\((\log p_i-\log p_j)/(p_i-p_j)\), with diagonal limit \(1/p_i\).
On a fixed commuting model Eq.~\eqref{eg:bkm_kernel} reduces to the
classical Fisher metric. Umegaki contrast selects this particular
quantum metric; monotonicity by itself permits a larger family
\cite{PetzSudar1996}.

At singular states relative entropy is finite only when
\(\operatorname{supp}\rho\subseteq\operatorname{supp}\sigma\).
Distinct pure states therefore have infinite relative entropy in either
direction. Their projective metric requires a separate construction;
a finite projector response does not guarantee a finite BKM boundary limit.

\begin{remark}
\label{eg:remark_connection_carrier}
Statistical metric connections act on tangent fibers; Berry connections
act on ray or spectral fibers. Identifying their transport requires an
explicit represented bundle map intertwining the connections
\cite{GauvinCompletion2026}. Appendix~\ref{app:geometry} gives the dual
connections, projective completion and general projector construction.
\end{remark}

\subsection{Fixed measurements and channel loss}
\label{sec:channels}
A finite positive-operator-valued measure (POVM) $\{M_a^x\}_a$ specifies the probabilities of outcomes $a$ at setting $x$. Its classical-record channel is
\begin{equation}
 \Phi_x(\rho)=\sum_a\Tr(\rho M_a^x)\ket a\bra a,
 \label{ch:measurement}
\end{equation}
where $M_a^x\ge0$ and $\sum_aM_a^x=\id$.
For a bipartite preparation,
\begin{equation}
 p(a,b|x,y)=\Tr[\rho_{AB}(M_a^x\otimes N_b^y)].
 \label{ch:born}
\end{equation}
Changing $x$ changes the readout map, whose Fisher pullback can have a kernel of invisible parameter directions.

Let $\Phi$ be a fixed completely positive trace-preserving (CPTP) map.
Evaluate output logarithms on $\mathcal K=\operatorname{ran}\Phi(\id)$,
the common faithful output support established in
Appendix~\ref{app:channel-details}.

Define the relative-entropy loss
\begin{equation}
 \mathcal L_\Phi(\rho,\sigma)
 =\DU(\rho\Vert\sigma)-\DU(\Phi\rho\Vert\Phi\sigma).
 \label{ch:loss}
\end{equation}
Data processing gives $\mathcal L_\Phi\ge0$~\cite{Lindblad1975}.
For trace-zero Hermitian density tangents $U,V$, its diagonal Hessian is
\begin{align}
 G_{\Phi,\rho}(U,V)
 &=\gBKM_\rho(U,V)
   -\gBKM_{\Phi\rho}(\Phi U,\Phi V),
 \label{ch:loss-metric}\\
 G_{\Phi,\rho}(U,U)&\ge0.
 \nonumber
\end{align}
Both tensors are pulled to the input carrier. The output metric of
Eq.~\eqref{ch:measurement} is Fisher information on active outcomes.
Appendix~\ref{app:channel-details} derives the Hessian and treats
composition, recovery, and changes of apparatus.

The coherence in Eq.~\eqref{ch:coherence} also satisfies
$C_r(\rho)=\DU(\rho\Vert\Delta\rho)/\ln2$~\cite{Baumgratz2014};
dephasing guarantees the required support inclusion.
For a faithful diagonal reference $\sigma=\Delta\sigma$, diagonal-log expectations also give
\begin{equation}
 \mathcal L_\Delta(\rho,\sigma)=\DU(\rho\Vert\Delta\rho).
 \label{ch:coherence-loss}
\end{equation}
Thus coherence in bits equals this channel loss divided by $\ln2$. Local unitaries preserve entanglement and optimized CHSH, but can change coherence in a fixed basis.

\subsection{A faithful noisy model}
\label{qm:sec:faithful}

For the Schmidt projector $P$ of Eq.~\eqref{qm:eq:state}, introduce depolarization:
\begin{equation}
 \rho_{p_{\mathrm{mix}}}=(1-p_{\mathrm{mix}})P+\frac{p_{\mathrm{mix}}}{4}\id,
 \qquad 0<p_{\mathrm{mix}}<1.
 \label{qm:eq:depolarized}
\end{equation}
Set $a=1-3p_{\mathrm{mix}}/4$, $b=p_{\mathrm{mix}}/4$, and $v=a-b=1-p_{\mathrm{mix}}$. The eigenvalue $a$ belongs to $P$ and $b$ to its orthogonal complement.

The following specialization of established monotone-metric kernels~\cite{PetzSudar1996} compares this noisy orbit with its unchanged eigenline geometry.

\begin{proposition}[Fixed-spectrum metric comparison]
\label{qm:prop:metric-comparison}
On the fixed-$p_{\mathrm{mix}}$ family in Eq.~\eqref{qm:eq:depolarized}, the BKM metric and symmetric logarithmic derivative (SLD) quantum Fisher information pull back to
\begin{align}
 \gBKM&=2v\ln(a/b)\,\gFS,\label{qm:eq:bkm}\\
 \gSLD&=\frac{4v^2}{a+b}\,\gFS.
 \label{qm:eq:sld}
\end{align}
The Bures metric is $\gSLD/4$ in these conventions.
\end{proposition}
The two-eigenvalue kernel calculation is given in Appendix~\ref{app:metric-comparison-proof}.

Although entropy is constant on this orbit, Eq.~\eqref{qm:eq:bkm} pulls back its ambient affine Hessian rather than differentiating the restricted constant. The nonlinear restriction need not inherit the ambient dually flat connections~\cite{GauvinCompletion2026}.

\subsection{Noise thresholds and boundary behavior}
\label{qm:subsec:noise}

The Pauli correlation matrix of Eq.~\eqref{qm:eq:depolarized} is $vT$. Thus its maximum over nondegenerate local spin observables is
\begin{equation}
 S_{\rm spin}=2v\sqrt{1+\sin^2(2\vartheta)}.
 \label{qm:eq:noisy-chsh}
\end{equation}
Allowing identity-valued measurements also permits the local value $2$. Bell violation occurs precisely when
\begin{equation}
 v>\frac1{\sqrt{1+\sin^2(2\vartheta)}}.
 \label{qm:eq:bell-threshold}
\end{equation}
The partial transpose has eigenvalues $b+vq$, $b+vp$, and $b\pm v\sqrt{pq}$. The necessary and sufficient two-qubit partial-transpose criterion~\cite{Peres1996,Horodecki1996} therefore gives entanglement exactly when
\begin{equation}
 v\sin(2\vartheta)>\frac{p_{\mathrm{mix}}}{2}.
 \label{qm:eq:ppt-threshold}
\end{equation}
At $\vartheta=\pi/4$, the state is entangled for $v>1/3$ but violates CHSH only for $v>1/\sqrt2$. Entanglement and positive geometric response can persist after CHSH violation is lost.

As $p_{\mathrm{mix}}\to0$, the SLD pullback tends to $4\gFS$ and the BKM coefficient diverges, consistent with the support restrictions in Sec.~\ref{eg:subsec_umegaki}. As $p_{\mathrm{mix}}\to1$, both pullbacks vanish at the constant state $\id/4$. The eigenline $P$ and its Berry phase persist for $p_{\mathrm{mix}}<1$ but lose their spectral identification at that degenerate endpoint. An operational mixed-state phase still requires an interference or purification protocol.

\begin{figure*}[t]
\centering
\begin{tikzpicture}
\begin{axis}[width=.47\textwidth,height=5.6cm,xmin=0,xmax=1,ymin=0,ymax=3.05,
 title={(a) Depolarization and Bell violation},title style={font=\small},
 xlabel={$p_{\mathrm{mix}}$},ylabel={$S_{\mathrm{spin}}$},
 xtick={0,.2928932188,.6666666667,1},xticklabels={$0$,$p_{\mathrm{mix},B}$,$p_{\mathrm{mix},E}$,$1$},
 tick label style={font=\small},label style={font=\small},
 legend style={font=\footnotesize,at={(.98,.97)},anchor=north east,draw=none},
 grid=major,grid style={gray!15}]
\addplot[blue!75!black,thick,domain=0:1,samples=2]{2*sqrt(2)*(1-x)};
\addlegendentry{Optimized spin CHSH}
\addplot[black,dashed,domain=0:1,samples=2]{2};
\addlegendentry{Local bound}
\draw[densely dashed,gray] (axis cs:.2928932188,0)--(axis cs:.2928932188,3.05);
\draw[densely dotted,gray] (axis cs:.6666666667,0)--(axis cs:.6666666667,3.05);
\end{axis}
\end{tikzpicture}\hfill
\begin{tikzpicture}
\begin{axis}[width=.47\textwidth,height=5.6cm,xmin=0,xmax=1,ymin=0,ymax=4,
 title={(b) Inequivalent entropy-metric limits},title style={font=\small},
 xlabel={$p_{\mathrm{mix}}$},ylabel={Metric factor $f$ in $g=4f\,\gFS$},
 tick label style={font=\small},label style={font=\small},
 legend style={font=\footnotesize,at={(.98,.97)},anchor=north east,draw=none},
 grid=major,grid style={gray!15}]
\addplot[red!75!black,thick] table[x=p_mix,y=bkm_scaled]{figures/depolarized.dat};
\addlegendentry{$f_{\mathrm{BKM}}=\tfrac v2\ln(a/b)$}
\addplot[blue!75!black,thick,dashed] table[x=p_mix,y=sld_scaled]{figures/depolarized.dat};
\addlegendentry{$f_{\mathrm{SLD}}=v^2/(a+b)$}
\addplot[blue!75!black,only marks,mark=o,mark size=1.8pt] coordinates {(0,1)};
\end{axis}
\end{tikzpicture}
\caption{Depolarized maximally entangled preparation, $\vartheta=\pi/4$. (a) CHSH violation ends at $p_{\mathrm{mix},B}=1-1/\sqrt2$ and entanglement at $p_{\mathrm{mix},E}=2/3$. Values below $2$ use nondegenerate spin observables. (b) BKM and SLD factors from Eqs.~\eqref{qm:eq:bkm} and~\eqref{qm:eq:sld}. The BKM curve starts at $p_{\mathrm{mix}}=.002$ and diverges at the pure-state boundary; the SLD factor tends to $1$ (open circle).}
\label{fig:noise}
\end{figure*}

\subsection{Flat statistical geometry with Bell correlations}
\label{lim:flat-comparison}
For a curvature counterexample, fix $P_\Phi=\ket{\Phi^+}\bra{\Phi^+}$ and consider the commuting family
\begin{align}
 \rho_v&=vP_\Phi+(1-v)\id_4/4,\quad 0<v<1,
 \label{lim:flat-family}\\
 \rho_\eta&=\frac{e^{\eta P_\Phi}}{e^\eta+3},\quad
 \eta=\ln\frac{1+3v}{1-v}.
 \label{lim:exponential}
\end{align}
Its eigenvalues $(1+3v)/4$ and $(1-v)/4$ (threefold) give
\begin{equation}
 \dd s^2=\frac{3\dd v^2}{(1-v)(1+3v)}.
 \label{lim:flat-metric}
\end{equation}
This one-dimensional Fisher/BKM manifold has zero intrinsic Riemann curvature, and its eigenline is fixed. Nevertheless, the two-qubit criteria give entanglement for $v>1/3$ and $S_{\mathrm{spin}}=2\sqrt2v>2$ for $v>1/\sqrt2$~\cite{Peres1996,Horodecki1996,Horodecki1995}, as in Sec.~\ref{qm:subsec:noise}. The fixed global eigenbasis is not a product basis. Spectral commutativity therefore does not imply locality relative to $A|B$.

Curvature is not sufficient either: the classical Fisher simplex is a spherical sector under $p_k\mapsto2\sqrt{p_k}$, with sectional curvature $1/4$ in dimension at least two~\cite{AmariNagaoka2000}. It supplies neither a quantum preparation nor a phase bundle. Metric components must also be distinguished from invariant curvature; the BKM boundary factors in Sec.~\ref{qm:sec:faithful} concern distinguishability and do not define a physical energy scale.

\subsection{A fixed measurement and the information it retains}
\label{qm:subsec:measurement}

For the computational-basis measurement, the pure family has probabilities $(q,0,0,p)$. On its fixed active support the Fisher metric is
\begin{equation}
 g^{\rm cl}_{\vartheta\vartheta}=4,
 \qquad g^{\rm cl}_{\phi\phi}=g^{\rm cl}_{\vartheta\phi}=0.
 \label{qm:eq:pure-fisher}
\end{equation}
For Eq.~\eqref{qm:eq:depolarized}, all four probabilities are positive:
\begin{equation}
 (q_0,q_1,q_2,q_3)=(b+vq,b,b,b+vp).
 \label{qm:eq:mixed-probabilities}
\end{equation}
Holding the measurement and $p_{\mathrm{mix}}$ fixed gives
\begin{align}
 g^{\rm cl}_{\vartheta\vartheta}
 &=v^2\sin^2(2\vartheta)
   \left(\frac1{b+vq}+\frac1{b+vp}\right),\nonumber\\
 g^{\rm cl}_{\phi\phi}&=g^{\rm cl}_{\vartheta\phi}=0.
 \label{qm:eq:mixed-fisher}
\end{align}
This readout erases phase and retains the full SLD $\vartheta$ information
at $\vartheta=\pi/4$. At the product endpoints with fixed
$p_{\mathrm{mix}}>0$, its first-order sensitivity vanishes despite a
nonzero quantum tangent. Thus the zero-noise and support-changing
endpoint limits do not commute.

\subsection{Imperfect records: metric retention and recoverable correlations}
\label{rec:sec}
We apply environment-assisted random-unitary correction
\cite{GregorattiWerner2003,TrendelkampSchroer2011} to a fixed imperfect
phase record. This single-copy protocol retains every trial and does not
optimize measurements on an inaccessible environment. The conditional
moments have an SLD antecedent in syndrome compression
\cite[Supplemental Theorem~S4, Eqs.~(S81)-(S83), and the noisy-readout
example in Eqs.~(S77)-(S78)]{Sheng2026Syndrome}. Here the retained
transverse entropy response yields sharp concurrence and CHSH intervals
for the specified two-qubit preparation.

Let the finite classical record $J$ have probabilities $w_j$, let
$V_j=\operatorname{diag}(1,e^{\ii\varphi_j})$ act locally on $B$, and obtain
$Y$ through a classical channel $p(y|j)$ with finite output alphabet. These probabilities and phases are
fixed independently of the input state. Restrict sums to nonzero-probability outcomes and
write
\begin{align}
 p_y&=\sum_jw_jp(y|j),&
 \mu_y&=\sum_jp(j|y)e^{\ii\varphi_j},\nonumber\\
 q_y&=|\mu_y|,&
 m&=\sum_yp_yq_y,\qquad
 \eta=\sum_yp_yq_y^2.
 \label{rec:moments}
\end{align}
Calibrated feedback $F_y=\operatorname{diag}(1,e^{-\ii\arg\mu_y})$
(arbitrary at $\mu_y=0$), followed by deleting $Y$, gives real coherence
multiplier $m$. The triangle inequality makes this optimal among
record-conditioned diagonal unitaries. Feedback precedes the independently
chosen Bell settings and uses neither postselection nor communication
during the test. Given the calibrated posterior moments, $m$ and the exact
resources in Eq.~\eqref{rec:resources} are directly computable. The
metric-only intervals instead use $\eta$ and the preparation parameters;
they quantify information-summary limits, not savings in measurement
or calibration cost.

The channel retaining the record is
\begin{equation}
 \mathcal F_Y(\rho)=\bigoplus_yp_y\Phi_y(\rho),\qquad
 \Phi_y(\rho)=\sum_jp(j|y)V_j\rho V_j^\dagger.
 \label{rec:flag}
\end{equation}
\begin{figure*}[t]
\centering
\begin{tikzpicture}
\begin{axis}[width=.47\textwidth,height=5.4cm,xmin=0,xmax=1,ymin=1.65,ymax=2.46,
 title={(a) Sharp metric-only interval},title style={font=\small},
 xlabel={Retained transverse metric $\eta$},ylabel={$S_{\rm spin}$},
 tick label style={font=\small},label style={font=\small},
 legend style={font=\footnotesize,at={(.03,.97)},anchor=north west,draw=none},
 grid=major,grid style={gray!15}]
\addplot[blue!75!black,thick,domain=0:1,samples=121]{1.7*sqrt(1+x*x)};
\addlegendentry{Erased sign record}
\addplot[teal!70!black,thick,dashed,domain=0:1,samples=121]{1.7*sqrt(1+x)};
\addlegendentry{Noisy sign record}
\addplot[black!65,densely dotted,domain=0:1]{2};
\addlegendentry{CHSH boundary}
\addplot[gray,densely dashed] coordinates {(.5,1.9006577809) (.5,2.0820662814)};
\addplot[only marks,mark=*,blue!75!black] coordinates {(.5,1.9006577809)};
\addplot[only marks,mark=square*,teal!70!black] coordinates {(.5,2.0820662814)};
\end{axis}
\end{tikzpicture}\hfill
\begin{tikzpicture}
\begin{axis}[width=.47\textwidth,height=5.4cm,xmin=.75,xmax=1,ymin=1.96,ymax=2.46,
 title={(b) Retaining the record versus compression},title style={font=\small},
 xlabel={Visibility $v$},ylabel={Optimal binary CHSH value},
 xtick={.75,.828427124746,.8944271909999,1},
 xticklabels={$0.75$,$v_{\rm keep}$,$v_{\rm qubit}$,$1$},
 ytick={2,2.1,2.2,2.3,2.4},
 tick label style={font=\small},label style={font=\small},
 legend style={font=\footnotesize,at={(.03,.97)},anchor=north west,draw=none,fill=white},
 grid=major,grid style={gray!15}]
\path[fill=orange!12] (axis cs:.828427124746,1.96)
 rectangle (axis cs:.8944271909999,2.46);
\draw[gray,densely dashed] (axis cs:.828427124746,1.96)--(axis cs:.828427124746,2.46);
\draw[gray,densely dashed] (axis cs:.8944271909999,1.96)--(axis cs:.8944271909999,2.46);
\addplot[blue!75!black,thick] coordinates
 {(.75,2) (.828427124746,2) (1,2.4142135623731)};
\addlegendentry{Retain Bob's record}
\addplot[teal!70!black,thick,dashed] coordinates
 {(.75,2) (.8944271909999,2) (1,2.2360679774998)};
\addlegendentry{Any qubit compression}
\addplot[black!65,densely dotted,domain=.75:1,samples=2]{2};
\addlegendentry{Binary-POVM floor}
\addplot[only marks,mark=o,mark size=2pt,black,mark options={fill=white},forget plot]
 coordinates {(.828427124746,2)};
\addplot[only marks,mark=*,mark size=2pt,black,forget plot]
 coordinates {(.8944271909999,2)};
\end{axis}
\end{tikzpicture}
\caption{Records of the same equiprobable phases $\pm\pi/2$.
(a) Sharp CHSH interval after optimal phase feedback and record deletion,
for $v=0.85$ and $\mathcal C_0=1$. Erased and noisy sign records attain
Eq.~\eqref{rec:resource-bounds}; at $\eta=1/2$, their opposite CHSH
outcomes survive full state error $0.01$. This panel uses nondegenerate
spin measurements.
(b) For the erased-sign record with $\eta=m=1/2$ and a depolarized Bell
input, the arbitrary-binary-POVM optima are
$S_{\rm keep}^{\star}=\max\{2,v(1+\sqrt2)\}$ and
$S_{\rm qubit}^{\star}=\max\{2,v\sqrt5\}$.
Shading marks $v_{\rm keep}<v\le v_{\rm qubit}$, where
$v_{\rm keep}=2/(1+\sqrt2)$ and $v_{\rm qubit}=2/\sqrt5$;
the open and filled markers indicate the excluded and included endpoints.
Only Bob holds the flag; Alice's measurements are common to all records.
Compression is any deterministic CPTP map $BY\to B'$ before the Bell
settings, with one output qubit and no accessible side output. All trials
are retained in both experiments.}
\label{fig:record-bounds}
\end{figure*}

\begin{proposition}[Sharp resource bounds from record retention]
\label{rec:prop}
For any faithful diagonal qubit reference $\sigma$ and nonzero transverse
Hermitian tangent $X=x_1\sigma_x+x_2\sigma_y$, the retained BKM metric fraction is
\begin{equation}
 \frac{\gBKM_{\mathcal F_Y(\sigma)}
       (\mathcal F_Y(X),\mathcal F_Y(X))}
      {\gBKM_\sigma(X,X)}=\eta.
 \label{rec:metric}
\end{equation}
The normalized entropy-loss Hessian is $1-\eta$: $\eta$ measures only
transverse retention, while diagonal tangents are unchanged. After optimal
feedback and record deletion, transverse retention is $m^2$, and
\begin{equation}
 \eta\le m\le\sqrt\eta,\qquad
 1-m^2=(1-\eta)+\operatorname{Var}_{p}(q_Y).
 \label{rec:bounds}
\end{equation}
The identity separates conditional phase uncertainty from variation in
record quality. Both bounds are sharp, and the same factors hold for every
normalized monotone metric at this reference. BKM supplies their
relative-entropy-loss interpretation, not a unique Bell characterization.

For the corrected local channel applied to the specified preparation
$\rho_v=vP+(1-v)\id_4/4$, $0\le v\le1$, with
$\mathcal C_0=\sin2\vartheta$ and the known initial phase aligned, direct two-qubit
criteria give
\begin{align}
 \mathcal C_{\rm out}&=\bigl[v\mathcal C_0m-(1-v)/2\bigr]_+,\nonumber\\
 S_{\rm spin}&=2v\sqrt{1+\mathcal C_0^2m^2},\qquad
 \mu_{\rm PT}=(1-v)/4-v\mathcal C_0m/2.
 \label{rec:resources}
\end{align}
Consequently $\eta$ alone gives the sharp intervals
\begin{align}
 \bigl[v\mathcal C_0\eta-(1-v)/2\bigr]_+
 &\le \mathcal C_{\rm out}\le
 \bigl[v\mathcal C_0\sqrt\eta-(1-v)/2\bigr]_+,\nonumber\\
 2v\sqrt{1+\mathcal C_0^2\eta^2}
 &\le S_{\rm spin}\le
 2v\sqrt{1+\mathcal C_0^2\eta}.
 \label{rec:resource-bounds}
\end{align}
Writing $\rho_{v,m}=v\rho_m+(1-v)\id_4/4$, with $\rho_m$ from
Eq.~\eqref{op:dephased-schmidt}, the relative entropy of coherence obeys
the sharp interval
\begin{equation}
 C_r(\rho_{v,\eta})\le C_r(\rho_{v,m})
                    \le C_r(\rho_{v,\sqrt\eta}).
 \label{rec:coherence-bounds}
\end{equation}
\end{proposition}
Appendix~\ref{rec:proof} gives the proof and equality conditions.
These bounds apply to the specified preparation and phase-noise class;
they do not define a metric witness for arbitrary states. For an independently justified full
trace-norm error $\epsilon$, a lower spin bound above $2+2\sqrt2\epsilon$
certifies CHSH violation; an upper bound below $2-2\sqrt2\epsilon$
excludes it for all binary measurements on the final unflagged qubits,
where $S_{\max}=\max\{2,S_{\rm spin}\}$. The exclusion covers neither other
Bell scenarios nor measurements retaining $Y$. Sufficient entanglement
and separability conditions are, respectively,
\begin{align}
 v\mathcal C_0\eta&>(1-v)/2+2\epsilon,\nonumber\\
 v\mathcal C_0\sqrt\eta&<(1-v)/2-2\epsilon,
 \label{rec:robust}
\end{align}
by Sec.~\ref{op:certification}. These strict margins leave equality cases
undecided and require an independent error budget including model and
record-calibration errors.

Take equiprobable $\varphi_j=\pm\pi/2$, realized by the finite loop at
$\beta=\pi/3$; discarding the phase label completely dephases $B$.
Revealing its sign with probability $\eta$ and otherwise reporting an
erasure gives $m=\eta$. A symmetric binary readout with error
$(1-\sqrt\eta)/2$ instead gives $m=\sqrt\eta$. The prior noise and retained
metric agree, but the recoverable correlations differ
(Fig.~\ref{fig:record-bounds}(a)). At $\eta=1/2$, $v=0.85$,
$\mathcal C_0=1$, the spin values are $1.90066$ and $2.08207$:
even with $\epsilon=0.01$, only the second output violates CHSH.

\emph{A combined control and recovery certificate.}
Implement these same phases with one circular loop at $\beta=\pi/3$ and
real fractional counterdiabatic error $|\zeta_j(t)|\le10^{-3}$.
Proposition~\ref{op:prop:quadratic}, followed by contraction under the fixed
classical readout, ideal phase feedback, and record discarding, gives
\begin{equation}
 \epsilon_{\rm ctrl}
 \le \tfrac12[10^{-3}\pi\sin(\pi/3)]^2
 <3.702\times10^{-6}.
 \label{rec:combined-error}
\end{equation}
At the same nominal $\eta=1/2$, $v=0.85$, $\mathcal C_0=1$, the CHSH
error margin is below $1.0471\times10^{-5}$, giving
$S_{\rm spin}^{\rm act}<1.90067$ for erased signs and
$S_{\rm spin}^{\rm act}>2.08205$ for noisy signs. This budget assumes
ideal preparation and feedback/readout, fixed record probabilities, and
the stated conditional error. Additional preparation, calibration, or
detector uncertainty must be included.

\subsection{Retaining a record versus compression into one qubit}
\label{rec:compression}
For the phase-aligned Bell preparation
$\rho_v=v\ket{\Phi^+}\bra{\Phi^+}+(1-v)\id_4/4$ ($\vartheta=\pi/4$),
compare two experiments. Only Bob holds $Y$; Alice uses her original
qubit and the same two measurements for every record. Bob either uses $Y$ to measure
$BY$, or applies any deterministic CPTP map
$\mathcal R:BY\to B'$ before the independent Bell settings, leaving
one qubit $B'$ and no other accessible output. Both experiments allow
arbitrary binary POVMs and retain every trial.

\begin{proposition}[Exact CHSH optima for a fixed record]
\label{rec:prop:compression}
For these two operation classes and the moments in
Eq.~\eqref{rec:moments},
\begin{align}
 S_{\rm qubit}^{\star}
 &=\max\{2,\,2v\sqrt{1+m^2}\},\nonumber\\
 S_{\rm keep}^{\star}
 &=\max\{2,\,2v\mathbb E\sqrt{1+q_Y^2}\}.
 \label{rec:compression-optima}
\end{align}
Conditional phase alignment attains the compression optimum over all
CPTP maps, extending its diagonal-unitary optimality for this preparation
and CHSH objective. If $q_Y$ is nonconstant, put
$F=\mathbb E\sqrt{1+q_Y^2}$ and $U=\sqrt{1+m^2}$.
Then $F>U$, and every visibility
\begin{equation}
 F^{-1}<v\le U^{-1}
 \label{rec:memory-window}
\end{equation}
gives a retained-record violation although no deterministic qubit
compression can violate CHSH.
\end{proposition}
Appendix~\ref{rec:compression-proof} proves both optimizations.
The spin branch of the retained optimum uses common Alice axes $\sigma_z,\sigma_x$ and,
after phase alignment, Bob axes
$(\sigma_z\pm q_y\sigma_x)/\sqrt{1+q_y^2}$.
Strict convexity of $\sqrt{1+q^2}$ gives Eq.~\eqref{rec:memory-window}.
The optimization excludes collective protocols, postselection,
setting-dependent compression and communication of $Y$ to Alice;
it is not an unrestricted LOCC bound.

Figure~\ref{fig:record-bounds}(b) shows the two optima for the erased-sign
record at $\eta=1/2$. At $v=0.85$,
\begin{equation}
 S_{\rm keep}^{\star}=0.85(1+\sqrt2)=2.05208\ldots,
 \qquad S_{\rm qubit}^{\star}=2.
 \label{rec:memory-example}
\end{equation}
The compressed spin ceiling is $1.90066\ldots$. A full trace-norm error
of at most $0.01$ on the flagged state leaves the retained witness
above $2.02379$ and every compressed spin value below $1.92895$.
CPTP contraction makes the latter bound uniform over all compressions.
Equation~\eqref{rec:combined-error} holds before record deletion and
supplies a smaller error under the same ideal preparation and readout
assumptions.

Measurement-relative compression can instead permit classical side
information~\cite{Bluhm2018Compression}. Any deterministic recovery can also be absorbed
into the final measurement through its adjoint
\cite[Sec.~II]{Scala2026RecoveryFree}. Our restriction is the one-qubit
output: direct measurements retain the larger $BY$ observable algebra.

\subsection{What finite-order entropy geometry can determine}
\label{rec:finite-geometry}
Adding local curvature need not make metric retention sufficient.
Pull back every normalized Petz monotone metric and the Umegaki contrast
along $\mathcal F_Y$ to the same affine Bloch input coordinates.
A metric jet through order $r$ comprises coefficient derivatives through
that order; a contrast jet includes mixed derivatives in both states.
Fixed record-dependent output unitaries preserve these pullbacks.
After phase alignment, $\Lambda_q:(x,y,z)\mapsto(qx,qy,z)$ has reliability
law $\nu$, with $M_k=\mathbb E_\nu q^{2k}$, $M_0=1$, $M_1=\eta$.

\begin{proposition}[Finite geometric data do not determine recovery]
\label{rec:prop:finite-jets}
For every integer $N\ge1$, there exist two finite readouts of the same
uniform phase prior $\varphi=\pm\pi/2$, with a common uniform flag
alphabet and identical diagonal output references, whose pulled-back
monotone metrics agree through order $2N-1$ at every faithful diagonal
reference and whose pulled-back Umegaki contrasts agree through total order $2N+1$
at every faithful diagonal pair. Yet for one common depolarized Bell
input, one optimized qubit-compression CHSH value exceeds $2$ and the
other equals $2$.
\end{proposition}
The statement ranges over finite records without a known cardinality
bound and concerns only the pulled-back input geometry, not the full
output state's intrinsic geometry or complete channel knowledge.

Appendix~\ref{rec:finite-proof} supplies the construction and proof.
For $N=2$ the two reliability laws are
\begin{align}
 \nu_- &: (q,p)=(0,1/4),\ (\sqrt{2/3},3/4),\nonumber\\
 \nu_+ &: (q,p)=(1/\sqrt3,3/4),\ (1,1/4).
 \label{rec:curvature-example}
\end{align}
Both have $M_1=1/2$ and $M_2=1/3$; splitting repeated outcomes gives
the same eight-element uniform flag alphabet. Metric derivatives through
order three agree, so the Riemann curvature tensors of all these
positive-definite pullback metrics agree at every faithful diagonal
reference. But $m_-=\sqrt6/4$ and $m_+=(1+\sqrt3)/4$.
At $v=0.84$ their optimal compressed spin values are $1.96997\ldots$
and $2.03447\ldots$. The opposite Bell outcomes survive a separately
justified full flagged trace-norm error $0.01$, uniformly over compressed
outputs on the nonviolating side. The loops and error model of
Eq.~\eqref{rec:combined-error} provide one such implementation budget.
The jet equalities describe the nominal channels; a state-accuracy
certificate does not certify experimental equality of channel derivatives.

Complete transverse response supplies the information absent from a
finite jet. For $\rho_t=(\id_2+t\sigma_x)/2$, $|t|<1$, define
\begin{align}
 H_\nu(t)&=\DU(\mathcal F_Y\rho_t\Vert\mathcal F_Y(\id_2/2))\nonumber\\
 &=\sum_{k\ge1}\frac{M_kt^{2k}}{2k(2k-1)}.
 \label{rec:complete-response}
\end{align}
Classical channel theory supplies the moment expansion
\cite[Sec.~IV B, Definition~16 and Lemmas~17 and~19]{Reeves2024Moments}
and reliability distributions~\cite{Goela2020Blackwell}. By
compact-interval moment uniqueness, all coefficients determine the law
of $q^2$, hence $\nu$, $m$, and both optima in
Eq.~\eqref{rec:compression-optima}. This does not guarantee stable
reconstruction from noisy finite data. If at most $K$ distinct reliabilities
are known to occur, $M_0,\ldots,M_{2K-1}$ suffice, corresponding to a
contrast jet through order $4K-2$. The required record size grows and its
Bell margin shrinks with $N$; no uniform high-order error tolerance is
asserted.

\section{Applications to physical transport and Bell thresholds}
\label{dyn:section}
\label{tr:sec}

The remaining sections apply the resource criteria to specified transport,
thermal, and detector channels. Lifetime predictions require a clock,
carrier representation, and evolution law. They concern crossings of the
attainable CHSH value; the operator bound $2\sqrt2$ remains unchanged.

Deterministic local transport preserves both entanglement and the
optimized value:
\begin{equation}
 \rho_t=(U_A\otimes U_B)\rho_0(U_A\otimes U_B)^\dagger.
 \label{dyn:local_unitary}
\end{equation}
Conjugating detector observables by their local unitaries preserves all
four correlators, including for independently represented spatial
translations. Fixed axes can give varying correlators; joint interactions
can change entanglement across the same partition.

\subsection{Proper time and correlated phase fluctuations}
\label{tr:clocks}

Local-unitary invariance is given by Eq.~\eqref{dyn:local_unitary};
here we specify the clock and phase statistics of the retained carriers.

For prescribed classical trajectories and fixed local energy bases, set
\begin{align}
 H_i&=E_{0i}\id+\Delta E_i\ket{1}\bra{1},\nonumber\\
 \phi_i(t)&=\hbar^{-1}\int_0^t\Delta E_i(s)\,\dd\tau_i(s),
 \label{tr:clock-phase}\\
 \frac{\dd\tau_i}{\dd t}
 &\simeq1+\frac{\Phi_g(\bm x_i(t))}{c_{\mathrm{light}}^2}
          -\frac{|\bm v_i(t)|^2}{2c_{\mathrm{light}}^2}.
 \label{tr:proper-time}
\end{align}
Equation~\eqref{tr:proper-time} retains leading weak-static-field and
slow-motion corrections: $\Phi_g$ has units $\mathrm{m^2\,s^{-2}}$,
$\Delta E_i$ is an energy, and $\phi_i$ is dimensionless
\cite{Zych2011,Pikovski2017}. For stationary equal-gap clocks,
$\phi_A-\phi_B\simeq\Delta E\,t\,\Delta\Phi_g/
(\hbar c_{\mathrm{light}}^2)$, with $\Delta\Phi_g\simeq g\Delta h$ near
a uniform field. An additive $f_i(t)\id$ changes only an overall phase,
whereas a level-gap change alters relative phase; a position-dependent
operator on superposed trajectories is not an identity shift.

Consider the initial states
\(\ket{\Phi_\vartheta}=\cos\vartheta\ket{00}+
\sin\vartheta\ket{11}\) and
\(\ket{\Psi_\vartheta}=\cos\vartheta\ket{01}+
\sin\vartheta\ket{10}\). Their accumulated relative phases are respectively
\(\phi_A+\phi_B\) and \(\phi_A-\phi_B\). Discarding a classical phase
record gives the characteristic functions
\begin{align}
 \kappa_\pm(t)&=\mathbb E e^{-\ii[\phi_A(t)\pm\phi_B(t)]},
 \label{tr:kappa}\\
 |\kappa_\pm(t)|
 &=\exp\!\left[-\frac{V_A(t)+V_B(t)\pm2C_{AB}(t)}2\right]
 \label{tr:gaussian}
\end{align}
for jointly Gaussian phases, where \(V_i=\operatorname{Var}\phi_i\)
and \(C_{AB}=\operatorname{Cov}(\phi_A,\phi_B)\).
If \(\delta\Omega_i(s)\) is the fluctuating angular frequency sampled
along trajectory \(i\), then
\begin{equation}
 C_{ij}(t)=\int_0^t\!\dd s\int_0^t\!\dd s'\,
  \mathbb E[\delta\Omega_i(s)\delta\Omega_j(s')],
 \qquad V_i=C_{ii}.
 \label{tr:spatial-covariance}
\end{equation}
The frequency covariance has units \(\mathrm{s^{-2}}\); its spatial
dependence supplies the separation dependence. Equal common phase noise
protects the \(\Psi\) subspace and enhances dephasing of the \(\Phi\)
subspace. Equal gaps, couplings, and timing are needed for this cancellation.

After a local basis exchange for the $\Psi$ family, the dephased-Schmidt
criterion, Eq.~\eqref{op:dephased-chsh}, gives
\begin{equation}
 S_{\mathrm{spin},\pm}(t)
 =2\sqrt{1+|\kappa_\pm(t)|^2\sin^2(2\vartheta)}.
 \label{tr:phase-chsh}
\end{equation}
Known mean phases rotate the optimal axes; an accessible record with
calibrated local control permits phase removal. Without the record,
finite Gaussian variances leave a nonzero CHSH excess at every finite
time when $\sin2\vartheta\ne0$. This commuting $U(1)$ model differs from
the physically represented, noncommuting $SU(2)$ diffusion of
Sec.~\ref{dyn:singlet_subsection}.

For a local Markov generator specified per unit proper time along a
deterministic trajectory, changing to laboratory time rescales its rate:
\begin{equation}
 \gamma_{\mathrm{lab},i}(t)=
 \gamma_{\mathrm{proper},i}(\tau_i(t))
 \frac{\dd\tau_i}{\dd t}.
 \label{tr:clock-rate}
\end{equation}
If the proper time fluctuates, the additional averaging can produce memory and need not preserve this Markov description.

\subsection{Represented fluctuations}

Choose a physical representation of transport and an unresolved-increment
law, as distinguished in Ref.~\cite{GauvinCompletion2026}. Assume independent
small Markov increments $\exp(-\ii T_a\dd\theta^a)$, with dimensionless
Hermitian $T_a$, zero leading mean, and covariance
$\mathbb E[\dd\theta^a\dd\theta^b]=D^{ab}\dd t$, where $D\succeq0$
has inverse-time units. Require higher-order increment contributions to
be $o(\dd t)$ in the diffusion limit. Expansion through order $\dd t$ gives
\begin{equation}
 \mathcal L(\rho)=-\tfrac12D^{ab}[T_a,[T_b,\rho]].
 \label{dyn:covariance_generator}
\end{equation}
A physical loop construction may supply
$\dd\theta^a=\mathcal F^a_I\dd\Sigma^I$, fixing the curvature-coefficient
convention for independent oriented area components $I$. With the same
diffusive scaling, the additional area-covariance postulate
\begin{equation}
 \mathbb E[\dd\Sigma^I\dd\Sigma^J]=C^{IJ}\dd t,\qquad
 D^{ab}=\mathcal F^a_I C^{IJ}\mathcal F^b_J
 \label{dyn:area_covariance}
\end{equation}
connects represented physical curvature to the channel. A statistical
connection alone supplies neither this representation nor the noise law;
an identity-valued phase has zero double commutator. Nonzero mean
increments or variable stochastic coefficients require the corresponding
drift terms. Ordinary Brownian coordinate increments are
$O(\sqrt{\dd t})$, but their small-loop stochastic areas are $O(\dd t)$,
with covariance $O((\dd t)^2)$. Equation~\eqref{dyn:area_covariance}
therefore requires an additional area-diffusion limit.
Section~\ref{op:finite-loop} gives a finite controlled-loop realization
without that limit.

For stationary independent isotropic increments $D^{ab}=\gamma\delta_{ab}$
and $\Tr(T_aT_b)=\delta_{ab}/2$, the defining representation of $SU(n)$
gives, on normalized states,
\begin{equation}
 \mathcal L(\rho)=\frac{\gamma n}{2}\left(\frac{\id_n}{n}-\rho\right).
 \label{dyn:su_n_generator}
\end{equation}
Thus $T_a=\sigma_a/2$ gives qubit Bloch contraction $e^{-\gamma t}$;
other representations can have multiple decay sectors.

\subsection{Correlated local rotations of a singlet}
\label{dyn:singlet_subsection}

On $\mathbb C^2\otimes\mathbb C^2$, set
$T_{Aa}=\sigma_a\otimes\id_2/2$ and
$T_{Ba}=\id_2\otimes\sigma_a/2$. Assume isotropic centered Gaussian angular increments, independent on disjoint time intervals,
with deterministic, locally integrable covariance coefficients
\begin{align}
 \mathbb E[\dd\theta_{ja}\dd\theta_{kb}]
 &=K_{jk}(t)\delta_{ab}\dd t,\nonumber\\
 K(t)&=\begin{pmatrix}\gamma_A(t)&c(t)\\c(t)&\gamma_B(t)\end{pmatrix}
 \succeq0.
 \label{dyn:local_covariance}
\end{align}
Thus $\gamma_j\ge0$ and $|c|\le\sqrt{\gamma_A\gamma_B}$ almost
 everywhere. Equation~\eqref{dyn:covariance_generator} becomes
\begin{align}
 \mathcal L_t(\rho)=-\tfrac12\sum_a\bigl(&
 \gamma_A[T_{Aa},[T_{Aa},\rho]]\nonumber\\
 &+\gamma_B[T_{Ba},[T_{Ba},\rho]]\nonumber\\
 &+2c[T_{Aa},[T_{Ba},\rho]]\bigr).
 \label{dyn:correlated_generator}
\end{align}
This CPTP-divisible evolution is homogeneous for constant coefficients.
Its finite-time channels are mixtures of product unitaries and preserve
separability, but their joint action depends on $c$, not merely the
marginal rates. The averaged state discards the rotation record; retaining
it would permit local inversion.

\begin{lemma}[Singlet decay under relative rotation noise]
\label{dyn:singlet_lemma}
For $P_-=\ket{\Psi^-}\bra{\Psi^-}$, where
$\ket{\Psi^-}=(\ket{01}-\ket{10})/\sqrt2$, the solution with
$\rho_0=P_-$ is
\begin{align}
 \rho_t&=v(t)P_-+[1-v(t)]\id_4/4,\nonumber\\
 v(t)&=e^{-\Gamma(t)},\qquad
 \Gamma(t)=\int_0^t R(s)\,\dd s,\label{dyn:werner_solution}\\
 R(t)&=\gamma_A(t)+\gamma_B(t)-2c(t)\ge0.
 \label{dyn:relative_rate}
\end{align}
\end{lemma}
\begin{proof}
Put $X=P_{-} -\id_4/4=-\tfrac14\sum_k\sigma_k\otimes\sigma_k$.
Pauli commutation gives
$\sum_a[T_{ja},[T_{ja},X]]=2X$ for each $j$.
The singlet commutes with every $T_{Aa}+T_{Ba}$, hence
$\sum_a[T_{Aa},[T_{Ba},X]]=-2X$.
Equation~\eqref{dyn:correlated_generator} consequently gives
$\mathcal L_t(X)=-R(t)X$ and annihilates $\id_4$.
The scalar differential equation proves the result; positivity of
$K$ gives $R\ge(\sqrt{\gamma_A}-\sqrt{\gamma_B})^2\ge0$.
\end{proof}

The correlation matrix is $-v\id_3$. Its optimal CHSH value and
concurrence are
\begin{equation}
 S_{\mathrm{spin}}=2\sqrt2e^{-\Gamma},\qquad
 \mathcal C=\left[\frac{3e^{-\Gamma}-1}{2}\right]_+,
 \label{dyn:resources}
\end{equation}
where $[x]_+=\max\{x,0\}$ \cite{Horodecki1995,Wootters1998}.
The two-qubit PPT criterion gives separability exactly when $v\le1/3$
\cite{Peres1996,Horodecki1996}. The resource crossings are therefore
\begin{equation}
 \Gamma_{\mathrm{CHSH}}=\ln\sqrt2,\qquad
 \Gamma_{\mathrm{ent}}=\ln3.
 \label{dyn:exponent_thresholds}
\end{equation}
For constant $R>0$, $t_X=\Gamma_X/R$; independent noise gives
\begin{equation}
 t_X=\frac{\Gamma_X}{\gamma_A+\gamma_B},\qquad
 X\in\{\mathrm{CHSH},\mathrm{ent}\}.
 \label{dyn:independent_times}
\end{equation}
The intervening entangled states do not violate CHSH. If the accumulated
exponent never reaches a threshold, that resource has no finite
crossing in this model. In particular, equal fully common noise has
$c=\gamma_A=\gamma_B$ and leaves the singlet invariant, as expected for
collective-noise protection \cite{Lidar1998}.

\subsection{A spatial covariance model}
\label{dyn:spatial_subsection}

To connect separation with relative noise, choose constant local rate
$\gamma>0$, correlation length $\ell_c>0$, and trajectory $L(t)=ut$ with
$u>0$. Postulate the positive spatial covariance
\begin{equation}
 c(L)=\gamma\exp[-L^2/(2\ell_c^2)].
 \label{dyn:gaussian_covariance}
\end{equation}
Here $c(L)$ is angular-increment cross covariance per unit time;
$c$ is not equal-time frequency covariance. The temporal white-noise
limit and spatial kernel are independent physical specifications.
Lemma~\ref{dyn:singlet_lemma} gives
\begin{equation}
 \Gamma(t)=2\gamma\left[t-\frac{\ell_c}{u}\sqrt{\frac\pi2}\,
 \operatorname{erf}\!\left(\frac{ut}{\sqrt2\ell_c}\right)\right].
 \label{dyn:gaussian_exponent}
\end{equation}
For these parameters it increases strictly for $t>0$ and is unbounded,
so each threshold in Eq.~\eqref{dyn:exponent_thresholds} has one finite
crossing. At small separation,
\begin{align}
 \Gamma(t)&=\frac{\gamma u^2t^3}{3\ell_c^2}
 -\frac{\gamma u^4t^5}{20\ell_c^4}+O(t^7),
 \label{dyn:early_exponent}\\
 t_X&\simeq
 \left(\frac{3\ell_c^2\Gamma_X}{\gamma u^2}\right)^{1/3}.
 \label{dyn:early_threshold}
\end{align}
This is the short-separation expansion after taking the white-noise limit. For a physical field with correlation time $\tau_c$, its use at a threshold additionally requires
\begin{equation}
 \tau_c\ll t_X\ll\ell_c/u,
 \qquad \gamma\tau_c\ll1,
 \label{dyn:memory-window}
\end{equation}
together with a valid coarse-grained generator. An equal-time spatial kernel alone does not supply these conditions.
At large separation the rate tends to $2\gamma$; at $u=0$ the continuous
limit gives exact singlet protection. The cubic exponent combines
quadratic spatial decorrelation with temporally white increments.
Figure~\ref{fig:dynamics-spatial} compares trajectories using
$a=\gamma\ell_c/u$.

\begin{figure*}[t]
\centering
\begin{tikzpicture}
\begin{axis}[width=.47\textwidth,height=5.7cm,xmode=log,xmin=.01,xmax=12,ymin=0,ymax=1.46,
 title={(a) CHSH response},title style={font=\small},
 xlabel={$\gamma t$},ylabel={$S_{\mathrm{spin}}/2$},
 xtick={.01,.1,1,10},xticklabels={$10^{-2}$,$10^{-1}$,$1$,$10$},
 tick label style={font=\small},label style={font=\small},
 legend style={font=\footnotesize,at={(.03,.04)},anchor=south west,draw=none,fill=white},
 grid=major,grid style={gray!15}]
\addplot[black,thick] table[x=gamma_t,y=S_ind]{figures/dynamics_spatial.dat};
\addlegendentry{Independent noise}
\addplot[blue!75!black,thick,dashed] table[x=gamma_t,y=S_a01]{figures/dynamics_spatial.dat};
\addlegendentry{$a=0.1$}
\addplot[red!75!black,thick,dashdotted] table[x=gamma_t,y=S_a1]{figures/dynamics_spatial.dat};
\addlegendentry{$a=1$}
\addplot[green!45!black,thick,densely dotted] table[x=gamma_t,y=S_a10]{figures/dynamics_spatial.dat};
\addlegendentry{$a=10$}
\addplot[gray,dashed,domain=.01:12,samples=2,forget plot]{1};
\end{axis}
\end{tikzpicture}\hfill
\begin{tikzpicture}
\begin{axis}[width=.47\textwidth,height=5.7cm,xmode=log,xmin=.01,xmax=12,ymin=0,ymax=1.04,
 title={(b) Entanglement survival},title style={font=\small},
 xlabel={$\gamma t$},ylabel={Concurrence},
 xtick={.01,.1,1,10},xticklabels={$10^{-2}$,$10^{-1}$,$1$,$10$},
 tick label style={font=\small},label style={font=\small},
 grid=major,grid style={gray!15}]
\addplot[black,thick] table[x=gamma_t,y=C_ind]{figures/dynamics_spatial.dat};
\addplot[blue!75!black,thick,dashed] table[x=gamma_t,y=C_a01]{figures/dynamics_spatial.dat};
\addplot[red!75!black,thick,dashdotted] table[x=gamma_t,y=C_a1]{figures/dynamics_spatial.dat};
\addplot[green!45!black,thick,densely dotted] table[x=gamma_t,y=C_a10]{figures/dynamics_spatial.dat};
\end{axis}
\end{tikzpicture}
\caption{Singlet evolution under local angular diffusion rate $\gamma$. Correlated curves use $c(L)=\gamma e^{-L^2/(2\ell_c^2)}$, $L=ut$, and $a=\gamma\ell_c/u$; independent noise has $c=0$. The dashed line in (a) marks the local CHSH bound. These white-noise predictions require the time-scale window in Eq.~\eqref{dyn:memory-window} for physical short-separation use; equal-time spatial covariance alone is insufficient.}
\label{fig:dynamics-spatial}
\end{figure*}

\subsection{A solvable finite-memory comparison}
\label{dyn:finite-memory}

A different microscopic short-time power can be exhibited without a Markov approximation. Consider commuting local $Z$ rotations with Hamiltonian $H_j(t)=\hbar\Omega(x_j(t),t)Z_j/2$, $x_A=0$, $x_B=ut$, and a linear spatial field $\Omega(x,t)=\Omega_0(t)+x\xi(t)$. The classical gradient is stationary centered Gaussian with
\begin{equation}
 \mathbb E[\xi(s)\xi(s')]=\sigma_\nabla^2 e^{-|s-s'|/\tau_c}.
 \label{dyn:gradient-covariance}
\end{equation}
The units of $\sigma_\nabla$ are $\mathrm{s^{-1}m^{-1}}$. The common term cancels for the $\ket{01},\ket{10}$ encoding. For this exactly specified linear field the relative phase is $u\int_0^t s\xi(s)\dd s$. Gaussian averaging, or the filter-function construction of Ref.~\cite{Cywinski2008}, gives
\begin{align}
 \kappa(t)&=\exp[-u^2\sigma_\nabla^2 J(t)/2],\nonumber\\
 J(t)&=\int_0^t\!\dd s\int_0^t\!\dd s'\,ss'e^{-|s-s'|/\tau_c}\nonumber\\
 &=\frac{2\tau_c t^3}{3}-\tau_c^2t^2
 +2\tau_c^4\bigl[1-(1+t/\tau_c)e^{-t/\tau_c}\bigr].
 \label{dyn:colored-integral}
\end{align}
Integrating over $s'\le s$ gives the last line and the limits
\begin{align}
 J(t)&=\frac{t^4}{4}-\frac{t^5}{15\tau_c}
       +O(t^6/\tau_c^2), &&t\ll\tau_c,\nonumber\\
 J(t)&=\frac{2\tau_ct^3}{3}
       \left[1-\frac{3\tau_c}{2t}
       +O((\tau_c/t)^3)\right], &&t\gg\tau_c.
 \label{dyn:colored-limits}
\end{align}
In particular $\chi=-\ln\kappa$ begins as $u^2\sigma_\nabla^2t^4/8$. The white-noise gradient limit holds $2\sigma_\nabla^2\tau_c$ fixed, and gives the cubic $\chi_{\mathrm W}=u^2\sigma_\nabla^2\tau_ct^3/3$. The exact integral satisfies, for $t>0$,
\begin{equation}
 0\le1-\frac{J(t)}{2\tau_ct^3/3}\le\frac{3\tau_c}{2t}.
 \label{dyn:colored-error}
\end{equation}
The upper inequality follows from the positive last term in Eq.~\eqref{dyn:colored-integral}; the lower follows by bounding the inner integral by $\tau_c s$. Thus the Markov exponent overestimates this finite-memory exponent by a controlled amount.

For an initial singlet this Abelian phase model gives $\mathcal C=\kappa$ and $S_{\mathrm{spin}}=2\sqrt{1+\kappa^2}$ (Sec.~\ref{tr:clocks}), both above their resource boundaries at finite $t$. It tests time scaling, not a finite isotropic-Werner death time. Approximating a general field by the linear model would also require bounds on omitted spatial derivatives.

The scale window in Eq.~\eqref{dyn:memory-window} is nonempty: choose
$\gamma\ell_c/u=10^6$ and $\gamma\tau_c=10^{-3}$. The cubic estimates give
$ut_B/\ell_c\simeq0.01013$ and $ut_E/\ell_c\simeq0.01488$, with
$t_B/\tau_c\simeq1.013\times10^7$. The fractional next spatial-series term,
$(3/20)(ut_X/\ell_c)^2$, is at most $3.4\times10^{-5}$ at these estimates.
These choices establish compatible scales;
Eq.~\eqref{dyn:colored-error} bounds only the separate commuting example,
not a noncommuting finite-memory singlet generator.

For the singlet trajectory, Appendix~\ref{dyn:bell_entropy_subsection}
computes a fixed-input relative-entropy distance to the Bell-local set,
which vanishes quadratically at a regular CHSH crossing with $R(t_B)>0$.
Appendix~\ref{dyn:mixing_subsection} gives a sufficient separability time
under entropy contraction toward a faithful product state, with an
explicit reset example.

\subsection{Retained carriers and detection}
\label{tr:detectors}

For distinguishable massive spin-$\tfrac12$ particles, a Lorentz
transformation is local on the full particle partition, but its
momentum-dependent Wigner rotations can change reduced spin correlations
\cite{Peres2002}. A fixed initially factorized momentum assignment gives
a CPTP spin map by unitary evolution and partial trace; initial
spin-momentum correlations require an assignment domain
\cite{Jordan2006}, and correlated momenta can induce correlated spin
noise. A passive frame change of both state and actual observables
preserves the probabilities of the same experiment.

Loss is a further channel, distinct from spin dephasing. For independent,
state-independent erasures with survival probabilities \(\eta_A,\eta_B\),
a setting-independent herald established before the measurement choices
defines a legitimate conditional experiment. Discarding no-click trials
after those choices does not automatically justify the same Bell bound.
For a maximally entangled pair, ideal CHSH axes, and the assignment \(+1\)
to every erasure, the all-trial expression is
\begin{equation}
 S_{\mathrm{all}}
 =2\sqrt2\,\eta_A\eta_B+2(1-\eta_A)(1-\eta_B).
 \label{tr:erasure-chsh}
\end{equation}
The surviving Bell marginal makes single-erasure terms vanish, while
double erasures contribute $2$.
For equal efficiencies this protocol violates CHSH precisely when
\(\eta>2/(1+\sqrt2)\), the conventional efficiency threshold
\cite{GargMermin1987}. The threshold is protocol-specific.

\section{Application to thermal storage and Bell lifetimes}
\label{th:sec:thermal}

Consider noninteracting stored qubits with independent stationary
reservoirs, initial state
$\rho_{AB}(0)\otimes\rho_{E_A}\otimes\rho_{E_B}$, and discarded reservoir
records. Weak coupling, short reservoir memory, resolved transition
frequencies, and the Born-Markov and secular approximations justify the
thermal semigroup limit \cite{Davies1974}. The following results are exact
for this reduced generator.

Kofman and Korotkov~\cite{KofmanKorotkov2008} analyzed relaxation,
dephasing, and distinct entanglement and CHSH survival times at arbitrary
temperature. Their principal preparation occupies $\ket{10},\ket{01}$,
whereas Eq.~\eqref{qm:eq:state} occupies $\ket{00},\ket{11}$.
Exchanging local energy levels changes the preparation's orientation
relative to a nonunital thermal channel; the specialization below retains
this dependence for the detector application.

\subsection{Generator, temperature, and rate conventions}
\label{th:subsec:generator}

Fix $\ket0$ as the ground state and $\ket1$ as the excited state:
\begin{align}
 H_j&=\hbar\omega_j\ket1\bra1,\qquad\omega_j>0,\nonumber\\
 \sigma_-&=\ket0\bra1,\qquad\sigma_+=\sigma_-^\dagger.
 \label{th:eq:energy-basis}
\end{align}
In the local interaction picture, let
\begin{align}
 \mathcal L_j(\rho)&=\Gamma_{\downarrow j}\mathcal D[\sigma_-]\rho
 +\Gamma_{\uparrow j}\mathcal D[\sigma_+]\rho\nonumber\\
 &\quad+\frac{\gamma_{\phi j}}2(\sigma_z\rho\sigma_z-\rho),\nonumber\\
 \mathcal D[L]\rho&=L\rho L^\dagger-\tfrac12\{L^\dagger L,\rho\}.
 \label{th:eq:generator}
\end{align}
All rates are constant and nonnegative, with $\Gamma_{1j}>0$. The population and off-diagonal decay rates are
\begin{align}
 T_{1j}^{-1}=\Gamma_{1j}&=\Gamma_{\downarrow j}+\Gamma_{\uparrow j},\nonumber\\
 T_{2j}^{-1}=\Gamma_{2j}&=\Gamma_{1j}/2+\gamma_{\phi j},\qquad
 r_j=\frac{\Gamma_{\uparrow j}}{\Gamma_{1j}}.
 \label{th:eq:rates}
\end{align}
Here $\gamma_{\phi j}$ is the additional off-diagonal decay rate, so its dissipator coefficient is halved; nonnegativity requires $T_{2j}\leq2T_{1j}$.

Write $\beta_j=(k_B T_{{\rm bath},j})^{-1}\geq0$, with zero temperature understood as $\beta_j\to\infty$ and infinite temperature as $\beta_j\to0$. Thermal detailed balance gives
\begin{equation}
 \frac{\Gamma_{\uparrow j}}{\Gamma_{\downarrow j}}
 =e^{-\beta_j\hbar\omega_j},\qquad
 r_j=\frac1{1+e^{\beta_j\hbar\omega_j}}.
 \label{th:eq:detailed-balance}
\end{equation}
For weak linear coupling to a bosonic reservoir, a standard parametrization is
\begin{align}
 \Gamma_{\downarrow j}&=\kappa_j(\omega_j)(n_j+1),&
 \Gamma_{\uparrow j}&=\kappa_j(\omega_j)n_j,\nonumber\\
 n_j&=(e^{\beta_j\hbar\omega_j}-1)^{-1}.
 \label{th:eq:bosonic-rates}
\end{align}
The spectral coupling $\kappa_j$ is physical input: temperature alone
does not fix $T_{1j}$. The energy basis fixes both coherence and
preparation orientation. Quasistatic broadening and echo depend on the
storage/readout protocol and need not be described by a constant
$\gamma_{\phi j}$; conditioning on a reservoir record, changing the
encoding, or active correction requires a different channel.

\subsection{Exact channel solution and correlation criteria}
\label{th:subsec:x-state}

Start from Eq.~\eqref{qm:eq:state}, with $p=\sin^2\vartheta$, $q=1-p$, and $\mathcal C_0=2\sqrt{pq}$. The range $0<p<1$ is retained: ordering Schmidt weights by exchanging energy levels would change their orientation relative to the dissipator. Define the single-site transition probabilities
\begin{align}
 \lambda_j&=e^{-\Gamma_{1j}t},&
 a_j&=r_j(1-\lambda_j),\nonumber\\
 b_j&=r_j+(1-r_j)\lambda_j.
 \label{th:eq:transition-probabilities}
\end{align}
Here $a_j$ is the excited population evolved from $\ket0$, and $b_j$ that evolved from $\ket1$. The independent channel
$\rho(t)=(e^{t\mathcal L_A}\otimes e^{t\mathcal L_B})\rho(0)$ yields
\begin{equation}
 \rho(t)=\begin{pmatrix}
 P_{00}&0&0&w\\
 0&P_{01}&0&0\\
 0&0&P_{10}&0\\
 w^*&0&0&P_{11}
 \end{pmatrix},
 \label{th:eq:x-state}
\end{equation}
where
\begin{align}
 P_{00}&=q(1-a_A)(1-a_B)+p(1-b_A)(1-b_B),\nonumber\\
 P_{01}&=q(1-a_A)a_B+p(1-b_A)b_B,\nonumber\\
 P_{10}&=qa_A(1-a_B)+pb_A(1-b_B),\nonumber\\
 P_{11}&=qa_Aa_B+pb_Ab_B,\nonumber\\
 w&=\sqrt{pq}\,e^{-(\Gamma_{2A}+\Gamma_{2B})t}e^{-\ii\phi}.
 \label{th:eq:x-entries}
\end{align}
The independent transition matrices evolve the populations, while each local $\ket0\bra1$ acquires $e^{-\Gamma_{2j}t}$. In the Schr\"odinger picture $\phi$ becomes $\phi-(\omega_A+\omega_B)t$, a local rotation covered by Eq.~\eqref{dyn:local_unitary}.

\begin{proposition}
\label{th:prop:x-resources}
For Eq.~\eqref{th:eq:x-state}, set
\begin{equation}
 k=2|w|,\qquad z=P_{00}+P_{11}-P_{01}-P_{10}.
 \label{th:eq:k-z}
\end{equation}
The concurrence and maximal spin CHSH value are
\begin{align}
 \mathcal C(t)&=2[|w|-\sqrt{P_{01}P_{10}}]_+,\nonumber\\
 S_{\rm spin}(t)&=2\sqrt{k^2+\max(k^2,z^2)},
 \label{th:eq:resources}
\end{align}
where $[x]_+=\max(0,x)$.
\end{proposition}
\begin{proof}
The two-qubit PPT criterion gives entanglement iff
$|w|^2>P_{01}P_{10}$ \cite{Peres1996,Horodecki1996}, and Wootters'
formula gives the stated concurrence \cite{Wootters1998}.
Local phase rotations make $w$ real, yielding
$T=\operatorname{diag}(k,-k,z)$; the Horodecki criterion then gives
$S_{\rm spin}$ \cite{Horodecki1995}.
\end{proof}

The measurement convention is that of Sec.~\ref{dyn:conventions}.
Relaxation and excitation populate $\ket{01}$ and $\ket{10}$, so the
computational-basis $l_1$ coherence $k$ implies entanglement only when
$k>2\sqrt{P_{01}P_{10}}$. At positive temperatures the limiting product
Gibbs state has strictly positive odd-sector populations while $w\to0$,
ensuring finite entanglement loss for this initially entangled family.
The diagonal marginals also contain thermal mixing, so their entropy is
no longer the pure-state entanglement entropy.

\subsection{Closed lifetimes and their preparation dependence}
\label{th:subsec:lifetimes}

For identical baths, write $\tau=t/T_1$, $\lambda=e^{-\tau}$, and $h=1-2r$. For the Bell preparation $\ket{\Phi^+}$ without additional pure dephasing,
\begin{align}
 k&=\lambda,& z&=\lambda^2+h^2(1-\lambda)^2,\nonumber\\
 P_{01}&=P_{10}=\frac{1-z}{4}.
 \label{th:eq:bell-thermal}
\end{align}
Define the loss time as the first crossing to $S_{\rm spin}\leq2$, and the entanglement-death time as the first crossing to $\mathcal C=0$. Then
\begin{align}
 t_{\rm CHSH}&=\frac{\ln2}{2}\,T_1,\nonumber\\
 t_{\rm ent}&=T_1\ln\!\left(1+\frac1{\sqrt{2r(1-r)}}\right),
 \qquad 0<r\leq\tfrac12.
 \label{th:eq:bell-lifetimes}
\end{align}
To obtain the first result, for $\lambda\geq1/\sqrt2$ note that
$z\leq\lambda^2+(1-\lambda)^2\leq\lambda$.
Hence $S_{\rm spin}=2\sqrt2\lambda$ until the crossing. Subsequent local
CPTP evolution cannot restore CHSH violation: its adjoints map binary
observables to admissible binary POVMs on the earlier state, whose maximum
is already 2. The spin-only value can still increase below 2.

The concurrence root $2\lambda=1-z$ has the unique positive solution
$\lambda_{\rm ent}=\sqrt R/(\sqrt2+\sqrt R)$, with $R=4r(1-r)$,
and occurs strictly after $t_{\rm CHSH}$.
At zero temperature, $\mathcal C(t)=\lambda^2$ and $t_{\rm ent}=\infty$.
Finite-temperature sudden death and its preparation dependence are
established reservoir effects \cite{Ali2009,YuEberly2004};
Eq.~\eqref{th:eq:bell-lifetimes} compares them with CHSH for this input.

\begin{table}[t]
\caption{Bell-state lifetimes for identical thermal baths without additional pure dephasing. Times are normalized by the actual relaxation time at the specified temperature. The last row is an infinite-temperature population limit within the model.}
\label{th:tab:lifetimes}
\begin{ruledtabular}
\begin{tabular}{ccc}
$r$ & $t_{\rm CHSH}/T_1$ & $t_{\rm ent}/T_1$\\
\hline
$0$ & $0.346574$ & $\infty$\\
$0.01$ & $0.346574$ & $2.092690$\\
$0.10$ & $0.346574$ & $1.211054$\\
$0.50$ & $0.346574$ & $0.881374$
\end{tabular}
\end{ruledtabular}
\end{table}

\begin{figure*}[t]
\centering
\begin{tikzpicture}
\begin{axis}[width=.47\textwidth,height=5.7cm,xmin=0,xmax=3,ymin=0,ymax=1.46,
 title={(a) Thermal CHSH response},title style={font=\small},
 xlabel={$t/T_1$},ylabel={$S_{\mathrm{spin}}/2$},
 tick label style={font=\small},label style={font=\small},
 legend columns=2,legend style={font=\footnotesize,at={(.97,.97)},anchor=north east,draw=none,fill=white},
 grid=major,grid style={gray!15}]
\addplot[black,thick] table[x=tau,y=S_r0]{figures/dynamics_thermal.dat};
\addlegendentry{$r=0$}
\addplot[blue!75!black,thick,dashed] table[x=tau,y=S_r001]{figures/dynamics_thermal.dat};
\addlegendentry{$r=0.01$}
\addplot[red!75!black,thick,dashdotted] table[x=tau,y=S_r01]{figures/dynamics_thermal.dat};
\addlegendentry{$r=0.1$}
\addplot[green!45!black,thick,densely dotted] table[x=tau,y=S_r05]{figures/dynamics_thermal.dat};
\addlegendentry{$r=0.5$}
\addplot[gray,dashed,domain=0:3,samples=2,forget plot]{1};
\draw[gray,densely dotted] (axis cs:.34657359028,0)--(axis cs:.34657359028,1.46);
\end{axis}
\end{tikzpicture}\hfill
\begin{tikzpicture}
\begin{axis}[width=.47\textwidth,height=5.7cm,xmin=0,xmax=3,ymin=0,ymax=1.04,
 title={(b) Thermal entanglement survival},title style={font=\small},
 xlabel={$t/T_1$},ylabel={Concurrence},
 tick label style={font=\small},label style={font=\small},
 grid=major,grid style={gray!15}]
\addplot[black,thick] table[x=tau,y=C_r0]{figures/dynamics_thermal.dat};
\addplot[blue!75!black,thick,dashed] table[x=tau,y=C_r001]{figures/dynamics_thermal.dat};
\addplot[red!75!black,thick,dashdotted] table[x=tau,y=C_r01]{figures/dynamics_thermal.dat};
\addplot[green!45!black,thick,densely dotted] table[x=tau,y=C_r05]{figures/dynamics_thermal.dat};
\end{axis}
\end{tikzpicture}
\caption{An initial $\lvert\Phi^+\rangle$ under identical independent thermal relaxation, with no additional pure dephasing. The equilibrium excited population is $r$; $r=0.5$ is the equal-rate limiting model. (a) The curves coincide up to the common CHSH crossing $t/T_1=(\ln2)/2$ (vertical line); the late increase at $r=0$ stays below the bound. (b) Concurrence has a nonzero finite-time tail at $r=0$ and disappears at finite time for $r>0$. Each curve uses its channel's actual $T_1$, generally temperature dependent.}
\label{fig:dynamics-thermal}
\end{figure*}

Figure~\ref{fig:dynamics-thermal} and Table~\ref{th:tab:lifetimes} use the actual $T_1(T)$; the formal high-temperature limit still requires weak coupling. Preparation orientation matters even at zero temperature: without added dephasing, the general Schmidt input has
\begin{equation}
 \mathcal C(t)=\lambda[\mathcal C_0-2p(1-\lambda)]_+.
 \label{th:eq:zero-temperature-general}
\end{equation}
For $0<p\leq1/2$ there is no finite entanglement-death time; for $p>1/2$ it is
\begin{equation}
 t_{\rm ent}=-T_1\ln\!\left(1-\sqrt{\frac{1-p}{p}}\right).
 \label{th:eq:orientation-lifetime}
\end{equation}
For $p=0.2$ and $p=0.8$, both with $\mathcal C_0=0.8$, the entanglement-death times are respectively infinite and $T_1\ln2$, while CHSH is lost at approximately $0.362551T_1$ and $0.123430T_1$. Exchanging populations changes robustness at fixed initial entanglement.

With identical additional pure dephasing, replace $k$ in Eq.~\eqref{th:eq:bell-thermal} by
\begin{equation}
 k=e^{-t/T_1-2\gamma_\phi t},
 \label{th:eq:added-dephasing}
\end{equation}
leaving $z$ unchanged. Equation~\eqref{th:eq:resources} gives the exact
roots: even at zero temperature, $d=\gamma_\phi T_1>0$ produces the
finite entanglement root $\lambda^{2d}=1-\lambda$.
For pure dephasing alone, the Bell-input special case of
Eq.~\eqref{op:dephased-chsh} instead has $\mathcal C=|\zeta|$ and
$S_{\rm spin}=2\sqrt{1+|\zeta|^2}$. Independent semigroups give
$|\zeta|=e^{-\nu t}$, $\nu=\gamma_{\phi A}+\gamma_{\phi B}$, so neither
resource is lost at finite time. A prescribed margin $S-2\geq\delta>0$
nevertheless requires
\begin{equation}
 t\leq\frac1{2\nu}\ln\!\frac1{\delta+\delta^2/4},
 \label{th:eq:certification-time}
\end{equation}
provided $\nu>0$ and the margin is initially attainable. This model
margin is not statistical confidence, which additionally requires a
sampling model, sufficient independent trials, and controlled detector
uncertainties; readout restrictions or conditioning must enter the
actual instrument (Sec.~\ref{tr:detectors}).

Calibrated $r_j,T_{1j},T_{2j}$ determine
$\Gamma_{\uparrow j}=r_j/T_{1j}$,
$\Gamma_{\downarrow j}=(1-r_j)/T_{1j}$, and
$\gamma_{\phi j}=T_{2j}^{-1}-(2T_{1j})^{-1}\ge0$; their uncertainties
propagate through the channel and threshold equations.
These preparation-dependent lifetimes are distinct from universal
channel entanglement breaking (Sec.~\ref{ur:subsec:eb}).

\subsection{An equilibrium reference that moves}
\label{tr:moving-gibbs}

The lifetime formulas assume fixed Hamiltonians and constant rates.
For a moving reference in finite dimension, let \(H_t\) be differentiable,
fix a finite inverse temperature \(\beta\), and define the faithful state
\(\tau_t=e^{-\beta H_t}/Z_t\). Assume
\begin{equation}
 \dot\rho_t=-\frac{\ii}{\hbar}[H_t,\rho_t]+\mathcal L_t(\rho_t),
 \qquad \mathcal L_t(\tau_t)=0,
 \label{tr:thermal-generator}
\end{equation}
where each frozen $\mathcal L_t$ is a
Gorini-Kossakowski-Lindblad-Sudarshan (GKLS) generator.
For faithful $\rho_t$,
frozen-semigroup contraction gives $\Sigma_t\ge0$ without detailed
balance \cite{Lindblad1975}, while differentiation of the reference gives
\begin{align}
 \Sigma_t&=-\Tr\!\left[
 \mathcal L_t(\rho_t)(\ln\rho_t-\ln\tau_t)\right],
 \label{tr:sigma}\\
 \frac{\dd}{\dd t}\DU(\rho_t\Vert\tau_t)
 &=-\Sigma_t+\beta\Tr[(\rho_t-\tau_t)\dot H_t].
 \label{tr:moving-reference}
\end{align}
Indeed, $\ln\tau_t=-\beta H_t-\ln Z_t$ and trace cyclicity gives
$\dd\ln Z_t/\dd t=-\beta\Tr(\tau_t\dot H_t)$ even for
$[H_t,\dot H_t]\ne0$; the coherent contribution vanishes because
$[H_t,\ln\tau_t]=0$. Singular states require a support-compatible
differentiable limit rather than direct use of the logarithmic
derivative at the boundary.

The work/reference term has either sign, so distance to a driven Gibbs
state need not decrease. An identity shift $H_t\mapsto H_t+f(t)\id$
leaves it unchanged because $\Tr(\rho_t-\tau_t)=0$.
For $\beta=\beta(t)$, add
$\dot\beta\,\Tr[(\rho_t-\tau_t)H_t]$; an extra coherent drive that does
not commute with $H_t$ and is omitted from $\mathcal L_t$ also requires
its own contribution to the derivative.

\section{Application to accelerated-detector channels}
\label{ur:sec:unruh}

Uniform acceleration changes a detector's channel through its field
response. Following Ref.~\cite{Tian2013}, we specialize the thermal
formulas and generalized-amplitude-damping criterion~\cite{LamiGiovannetti2015}
to a specified response, then quantify the accuracy needed to certify
their boundaries.

\subsection{A local detector response and its assumptions}
\label{ur:subsec:response}

Consider a two-level Unruh-DeWitt detector with fixed proper gap $\Delta E=\hbar\Omega>0$ and constant linear proper acceleration $a$ in the Minkowski vacuum. Its stationary temperature~\cite{Unruh1976}, comparison scale, and excited-state population are
\begin{align}
 k_B T_U&=\frac{\hbar a}{2\pi c_{\mathrm{light}}},
 \label{ur:eq:temperature}\\
 a_\Delta&=2\pi c_{\mathrm{light}}\Omega,\qquad
 \alpha=\frac{a}{a_\Delta},
 \label{ur:eq:acceleration-scale}\\
 x&=\frac{\pi c_{\mathrm{light}}\Omega}{a}=\frac1{2\alpha},
 \qquad r=\frac1{1+e^{2x}}.
 \label{ur:eq:occupation}
\end{align}
The scale $a_\Delta$ equates gap and thermal energy; it is not a resource threshold. The population $r$ is a two-level Gibbs probability, irrespective of field statistics. Detailed balance fixes the rate ratio, not its magnitude.

Use a pointlike detector with transverse monopole coupling to a massless scalar field in $3+1$ dimensions. In the stationary weak-coupling Markov and secular description, its inertial vacuum spontaneous-emission rate $\Gamma_0>0$ fixes the local response~\cite{HuYu2015,Moustos2018}:
\begin{align}
 \Gamma_\downarrow&=\frac{\Gamma_0}{1-e^{-2x}},&
 \Gamma_\uparrow&=\frac{\Gamma_0}{e^{2x}-1},\nonumber\\
 \Gamma_1&=\Gamma_0\coth x,& T_1&=\Gamma_1^{-1}.
 \label{ur:eq:rates}
\end{align}
These rates use proper time $\tau$. With no added pure dephasing,
$\Gamma_2=\Gamma_1/2$; the interaction-picture generator is
Eq.~\eqref{th:eq:generator} with $\gamma_\phi=0$ and the energy labels of
Eq.~\eqref{th:eq:energy-basis}. Each inversion fixes the calibrated proper
gap and acceleration-independent $\Gamma_0$. Tracking known Hamiltonian
phases changes neither concurrence nor optimized CHSH. Switching and
memory corrections are neglected; physical gap or coupling-spectrum
changes alter the rates~\cite{Moustos2018}.

For two detectors, use independent scalar-field copies in their Minkowski
vacua, each coupled only to its own detector. The initial state is
$\rho_{AB}(0)\otimes\ket{0_A}\bra{0_A}\otimes\ket{0_B}\bra{0_B}$.
Equal gaps, accelerations, and proper exposures give the product reduced
channel with no retained field records. Independence is exact for these
copies. A common field instead requires bounds on cross dissipation and
coherent coupling~\cite{HuYu2015}. Retained records or corrective
instruments define a different operational problem.

First form the stationary weak-coupling generator at fixed $a>0$ and
$\Omega>0$, holding $\Gamma_0\tau$ finite as coupling decreases~\cite{Davies1974}.
The subsequent limits $a\to0$ and $\tau\to\infty$ belong to that semigroup.
No finite-coupling approximation bound uniform in either limit is derived:
exact zeros, infinite tails, and sharp acceleration boundaries remain
semigroup properties pending an independent error estimate resolving their
resource margins. At zero coupling the dissipative channel is absent;
formulas with $\Gamma_0>0$ have only a limiting interpretation.

\subsection{Two noisy detectors: exact state-dependent thresholds}
\label{ur:subsec:pair}

Prepare $\ket{\Phi^+}=(\ket{00}+\ket{11})/\sqrt2$ and apply the channel independently to both halves. Set
\begin{equation}
 \lambda=e^{-\Gamma_1\tau},\qquad
 z=\lambda^2+\tanh^2x\,(1-\lambda)^2.
 \label{ur:eq:pair-parameters}
\end{equation}
Equation~\eqref{th:eq:resources} yields
\begin{align}
 S_{\mathrm{spin},2}&=2\sqrt{\lambda^2+\max(\lambda^2,z^2)},\nonumber\\
 \mathcal C_2&=\left[\lambda-\frac{1-z}{2}\right]_+
 =\left[\lambda^2-2r(1-r)(1-\lambda)^2\right]_+.
 \label{ur:eq:pair-resources}
\end{align}
The suffix 2 denotes two noisy detectors. Joint computational-basis $l_1$ coherence is $\lambda$ and can survive the concurrence zero (Sec.~\ref{th:subsec:x-state}).

The single-copy, all-outcome convention of Sec.~\ref{dyn:conventions} applies: optimized binary POVMs give $\max(2,S_{\mathrm{spin},2})$, and CHSH nonviolation does not establish locality for every Bell protocol.

Substituting Eq.~\eqref{ur:eq:rates} into Eq.~\eqref{th:eq:bell-lifetimes} gives
\begin{align}
 \tau_{B,2}(a)&=\frac{\ln2}{2\Gamma_0}\tanh x,\nonumber\\
 \tau_{E,2}(a)&=\frac{\tanh x}{\Gamma_0}
                   \ln(1+\sqrt2\cosh x).
 \label{ur:eq:pair-lifetimes}
\end{align}
For $a>0$, this preparation violates CHSH exactly for $0\leq\tau<\tau_{B,2}$ and is entangled exactly for $0\leq\tau<\tau_{E,2}$; equality excludes the corresponding strict resource. The concurrence crossing follows the Bell crossing. Physical certification requires the margins of Sec.~\ref{op:certification} and a detector error budget, as discussed in Sec.~\ref{ur:subsec:robustness}.

At $a=0$, understood as a limit, vacuum relaxation persists: $\Gamma_1=\Gamma_0$, $\mathcal C_2=e^{-2\Gamma_0\tau}$ has no finite zero, and $\tau_{B,2}=\ln2/(2\Gamma_0)$. An inertial trajectory does not remove spontaneous emission from the chosen open-system model.

\subsection{Acceleration bounds at fixed proper exposure}
\label{ur:subsec:inversion}

Let $s=\Gamma_0\tau>0$ be a required exposure at fixed $\Omega,\Gamma_0$. A positive CHSH acceleration boundary exists only for $s<\ln2/2$:
\begin{equation}
 a_{B,2}(\tau)=\frac{\pi c_{\mathrm{light}}\Omega}
 {\operatorname{artanh}(2\Gamma_0\tau/\ln2)},
 \qquad 0<s<\tfrac12\ln2.
 \label{ur:eq:bell-acceleration}
\end{equation}
Strict violation survives for $0\leq a<a_{B,2}$. For $s\geq\ln2/2$, no nonnegative acceleration retains it. At zero exposure the ideal input remains a Bell state for every finite acceleration.

For entanglement, define
\begin{equation}
 f(x)=\tanh x\ln(1+\sqrt2\cosh x).
 \label{ur:eq:inverse-function}
\end{equation}
Both factors are positive and strictly increasing for $x>0$. Since $f(0)=0$ and $f(x)\to\infty$, each $s>0$ has a unique positive root $f(x_E)=s$, giving
\begin{equation}
 a_{E,2}(\tau)=\frac{\pi c_{\mathrm{light}}\Omega}{x_E},
 \qquad f(x_E)=\Gamma_0\tau.
 \label{ur:eq:entanglement-acceleration}
\end{equation}
Concurrence is positive exactly for $0\leq a<a_{E,2}$, with $a_{E,2}>a_{B,2}$ whenever the Bell boundary exists. Both lifetimes decrease strictly with acceleration. Figure~\ref{fig:unruh} compares them with the channel threshold below; Sec.~\ref{uc:thresholds} gives a normalized numerical example.

Formally, at large $a$, $\Gamma_0\tau_{B,2}\sim x\ln2/2$ and $\Gamma_0\tau_{E,2}\sim x\ln(1+\sqrt2)$. As $a\to0^+$, $\Gamma_0\tau_{E,2}=x-\ln\sqrt2+o(1)$.

\begin{figure*}[t]
\centering
\begin{tikzpicture}
\begin{axis}[width=.47\textwidth,height=6.1cm,xmode=log,ymode=log,
 xmin=.01,xmax=100,ymin=.001,ymax=120,
 title={(a) Proper exposure thresholds},title style={font=\small},
 xlabel={$a/a_\Delta$},ylabel={$\Gamma_0\tau$},
 tick label style={font=\small},label style={font=\small},
 legend style={font=\footnotesize,at={(.98,.98)},anchor=north east,draw=none,fill=white},
 grid=major,grid style={gray!15}]
\addplot[blue!75!black,thick] table[x=alpha,y=bell]{figures/unruh_dynamics.dat};
\addlegendentry{Two-site CHSH}
\addplot[red!75!black,thick,dashed] table[x=alpha,y=ent]{figures/unruh_dynamics.dat};
\addlegendentry{Two-site entanglement}
\addplot[green!40!black,thick,dashdotted] table[x=alpha,y=eb]{figures/unruh_dynamics.dat};
\addlegendentry{One-site channel EB}
\end{axis}
\end{tikzpicture}\hfill
\begin{tikzpicture}
\begin{axis}[width=.47\textwidth,height=6.1cm,xmode=log,
 xmin=.1,xmax=10,ymin=0,ymax=1.48,
 title={(b) Fixed exposure $\Gamma_0\tau=0.1$},title style={font=\small},
 xlabel={$a/a_\Delta$},ylabel={Resource value},
 tick label style={font=\small},label style={font=\small},
 legend style={font=\footnotesize,at={(.98,.98)},anchor=north east,draw=none,fill=white},
 grid=major,grid style={gray!15}]

\addplot[blue!75!black,thick] table[x=alpha,y=S_half]{figures/unruh_dynamics.dat};
\addlegendentry{$S_{\rm spin}/2$}
\addplot[red!75!black,thick,dashed] table[x=alpha,y=C]{figures/unruh_dynamics.dat};
\addlegendentry{Concurrence}
\addplot[gray,dashed,domain=.1:10,samples=2,forget plot]{1};
\draw[gray,densely dotted] (axis cs:1.68366122605,0)--(axis cs:1.68366122605,1.48);
\draw[gray,densely dotted] (axis cs:4.406811292,0)--(axis cs:4.406811292,1.48);
\end{axis}
\end{tikzpicture}
\caption{Independent scalar-detector predictions, with $a_\Delta=2\pi c_{\mathrm{light}}\Omega$ and $x=1/[2(a/a_\Delta)]$. (a) Two identical channels act on $\lvert\Phi^+\rangle$ for the CHSH and entanglement curves; the one-site curve instead marks channel entanglement breaking (EB) for every input and reference. (b) Both sites have proper exposure $\Gamma_0\tau=0.1$. The shaded interval $1.68366<a/a_\Delta<4.40681$ retains entanglement without CHSH violation. These are ideal semigroup boundaries; certification requires the error margins in Sec.~\ref{op:certification}. Table~\ref{ur:tab:robustness} gives a specified perturbation example.}
\label{fig:unruh}
\end{figure*}

\subsection{One noisy detector and an exact channel threshold}
\label{ur:subsec:eb}

Isolate one member of the Bell pair while the other experiences $\mathcal E_\tau$. Isolation means identity evolution up to a known local unitary, not an inertial trajectory with continuing field coupling. The one-sided output has
\begin{align}
 S_{\mathrm{spin},1}&=2\sqrt{2\lambda},\nonumber\\
 \mathcal C_1&=\left[\sqrt\lambda-
             \sqrt{r(1-r)}(1-\lambda)\right]_+,
 \label{ur:eq:one-sided-resources}
\end{align}
and $\tau_{B,1}=(\ln2)\tanh x/\Gamma_0=2\tau_{B,2}$.

This one-sided output is the Choi state $J(\mathcal E_\tau)$ of Eq.~\eqref{ur:eq:choi}; its separability tests entanglement breaking. The known generalized-amplitude-damping criterion~\cite[Eq.~(27)]{LamiGiovannetti2015}, with parameters $p=\lambda$ and $\gamma=1-r$, is equivalent to
\begin{equation}
 \lambda\leq r(1-r)(1-\lambda)^2.
 \label{ur:eq:eb-condition}
\end{equation}
Its boundary is $\lambda_*=e^{-2\operatorname{arsinh}(\cosh x)}$. Substituting the scalar rate gives, for $a>0$, entanglement breaking exactly when $\tau\geq\tau_{\mathrm{EB}}$, where
\begin{equation}
 \tau_{\mathrm{EB}}(a)=\frac{2\tanh x}{\Gamma_0}
                       \operatorname{arsinh}(\cosh x).
 \label{ur:eq:eb-time}
\end{equation}
Single-site entanglement breaking is distinct from two-site entanglement annihilation~\cite{Filippov2012}. It concerns every input and reference; the earlier $\tau_{E,2}$ concerns the Bell input alone. Entanglement breaking need not remove basis coherence. At $a=0$ there is no finite entanglement-breaking time and $\mathcal C_1=e^{-\Gamma_0\tau/2}$.

At fixed $s$, $g(x)=2\tanh x\operatorname{arsinh}(\cosh x)$ increases strictly from zero to infinity. Thus the channel is entanglement breaking exactly for $a\geq\pi c_{\mathrm{light}}\Omega/x_{\mathrm{EB}}$, where $g(x_{\mathrm{EB}})=s$.

\subsection{Low-temperature sensitivity and a controlled perturbation}
\label{ur:subsec:robustness}

The detector application now uses the margins of Sec.~\ref{op:certification}.

The low-temperature root $\lambda_E=\sqrt{2r(1-r)}/[1+\sqrt{2r(1-r)}]$ is especially sensitive. For the two-channel Bell output, the potentially negative partial-transpose eigenvalue obeys
\begin{align}
 m(\tau)&=r(1-r)(1-\lambda)^2-\lambda^2/2,\nonumber\\
 m'(\tau_{E,2})&=\Gamma_1\frac{\lambda_E^2}{1-\lambda_E}
 \sim2\Gamma_1r\quad(r\to0).
 \label{ur:eq:ppt-sensitivity}
\end{align}
This is the smallest eigenvalue near the crossing. At $\alpha=0.01$,
$r\simeq3.72\times10^{-44}$ and $\Gamma_0\tau_{E,2}\simeq49.65$; the
equilibrium minimum is $r^2$, not $r(1-r)$. A small absolute state error
can leave the death time unresolved despite an accurate decay curve.
Perturbative state accuracy and low-temperature entanglement require
separate control~\cite{FlemingCummings2011,FlemingEtAl2012}; the semigroup
alone does not establish finite-coupling zero-temperature tails or exact
physical death times.

\begin{table}[t]
\caption{Certificates at $\Gamma_0\tau=0.1$ for the perturbation in Eq.~\eqref{ur:eq:perturbation-example}. Here $\alpha=a/a_\Delta$, $\mu_2$ is the model two-site PPT minimum, and $\mu_J$ is the model one-site Choi PPT minimum. The CHSH margin is $2\sqrt2\epsilon_2\simeq0.0028284$; the PPT margins are $\epsilon_2=0.001$ and $\epsilon_J=0.0005$.}
\label{ur:tab:robustness}
\begin{ruledtabular}
\begin{tabular}{ccrl}
Quantity & $\alpha$ & Model value & Certificate\\
\hline
$S_{\rm spin}-2$ & $1.6$ & $0.032713$ & Violation\\
$S_{\rm spin}-2$ & $1.8$ & $-0.044768$ & No violation\\
$\mu_2$ & $4.3$ & $-0.0063054$ & Entangled\\
$\mu_2$ & $4.5$ & $0.0053626$ & Separable\\
$\mu_J$ & $8.7$ & $-0.0034059$ & Not EB\\
$\mu_J$ & $8.9$ & $0.0024520$ & EB
\end{tabular}
\end{ruledtabular}
\end{table}

For a controlled illustration, suppose $\mathcal K_u$ and $\widetilde{\mathcal K}_u$ both generate CPTP propagators and their difference has diamond norm at most $\delta(u)$. The diamond norm is the induced trace norm stabilized by an arbitrary reference. Duhamel's formula and the unit diamond norm of a channel give
\begin{equation}
 \|\widetilde{\mathcal E}_{\tau,0}-\mathcal E_{\tau,0}\|_\diamond
 \le\int_0^\tau\delta(u)\,\dd u.
 \label{ur:eq:generator-error}
\end{equation}
An initial preparation error can be added to this bound by contractivity.

For example, retain the same ideal preparation and readout and the independent scalar channels, but allow residual local dephasing
\begin{equation}
 \widetilde{\mathcal L}_j=\mathcal L_j+\kappa_j\Gamma_0
 \mathcal D[\sigma_z],\qquad 0\le\kappa_j\le2.5\times10^{-3}.
 \label{ur:eq:perturbation-example}
\end{equation}
Here $\mathcal D[\sigma_z]$ has diamond norm two, attained on $\ket{+}$, as the difference of a unitary channel and the identity. At $s=\Gamma_0\tau=0.1$ Eq.~\eqref{ur:eq:generator-error} gives $\epsilon_2=10^{-3}$ for the two-site output and $\epsilon_J=5\times10^{-4}$ for a one-site Choi state. The two-site coherence is multiplied by $e^{-2(\kappa_A+\kappa_B)s}$ and its exact trace-norm change is $\lambda[1-e^{-2(\kappa_A+\kappa_B)s}]\le\epsilon_2$. The norm certificates in Table~\ref{ur:tab:robustness} hold throughout this dephasing range. These budgets describe specified generator uncertainty; microscopic switching, memory, and weak-coupling errors must be separately bounded and added for experimental certification.

\subsection{Clocks, geometric response, and physical scope}
\label{ur:subsec:scope}

The inversions fix detector proper exposure. Uniform acceleration starting at rest in a chosen inertial frame relates it to laboratory time by
\begin{equation}
 \tau=\frac{c_{\mathrm{light}}}{a}
       \operatorname{arsinh}\!\left(\frac{a t_{\mathrm{lab}}}
                                        {c_{\mathrm{light}}}\right).
 \label{ur:eq:clock-conversion}
\end{equation}
At fixed laboratory duration this expression must enter the channel before solving the boundary; monotonicity under one clock protocol does not establish it under another.

Writing $\eta_0=\Gamma_0/\Omega$, the scalar secular condition is
\begin{equation}
 \eta_0\coth[1/(2\alpha)]\ll1,
 \label{ur:eq:secular-window}
\end{equation}
or $2\eta_0\alpha\ll1$ at large $\alpha$. For $\eta_0=10^{-4}$ and $s=0.1$, $\Omega\tau=10^3$. Across $1.6\le\alpha\le8.9$, covering Table~\ref{ur:tab:robustness}, $\Gamma_1/\Omega\le1.782\times10^{-3}$. This checks a scale hierarchy, not a microscopic error bound.

The same clock relation gives rapidity
$a\tau/c_{\mathrm{light}}=2\pi\alpha(\Omega\tau)$ and terminal Lorentz
factor $\gamma_{\mathrm L}=\cosh(a\tau/c_{\mathrm{light}})$ in the chosen
initial rest frame. Already at $\alpha=1.6$ and $\Omega\tau=10^3$, the
rapidity is $1.0053\times10^4$ and $\log_{10}\gamma_{\mathrm L}\simeq4365.7$.
These are abstract semigroup hierarchy examples. Establishing an
attainable accelerator protocol additionally requires a device model for
the worldline duration and energy budget.

A physical protocol must specify its worldline and switching
envelope~\cite{LoukoSatz2008,FewsterJuarezLouko2016}, match preparation to
the factorized input, and budget switch-off error. Optimized readout
requires calibrated local rotations, binary measurements, and known
phase and timing records. Coupling during these operations enters the
exposure/readout channel; erasures follow Sec.~\ref{tr:detectors}.
A stationary KMS ratio does not bound finite-duration errors; no numerical
switching-error bound is assigned here.

For $a>0$, the commuting stationary family $\sigma_a=\operatorname{diag}(1-r_a,r_a)$ has BKM (Fisher) metric
\begin{equation}
 g_{aa}=\frac{(\partial_a r_a)^2}{r_a(1-r_a)}
        =\frac{a_\Delta^2}{a^4}r_a(1-r_a).
 \label{ur:eq:equilibrium-metric}
\end{equation}
This stationary metric contains no relaxation rate. Its one-dimensional
intrinsic curvature vanishes, and fixed energy eigenvectors supply no
Berry curvature. Known local Hamiltonian loops preserve entanglement and
optimized CHSH; the losses arise from the specified interaction and
discarded field information.

\section{Carrier scales for the detector applications}
\label{uc:sec}

\subsection{The transition gap sets the thermal comparison}
\label{uc:scale}

Comparing carriers requires their encoded levels and coupling. For a positive gap $\Delta E=\hbar\Omega=h\nu$, the scale in Eq.~\eqref{ur:eq:acceleration-scale} becomes
\begin{equation}
 a_\Delta=4\pi^2c_{\mathrm{light}}\nu.
 \label{uc:acceleration-scale}
\end{equation}
This scale equates gap and thermal energy. The Gibbs population in
Eq.~\eqref{ur:eq:occupation} describes equilibrium, not a loss threshold
or finite-time excitation probability. Small gaps can increase thermal
sensitivity while weak coupling makes equilibration slow.

Using the SI defining constants gives
\begin{align}
 T_U&=(4.05501\times10^{-21}\,
       \mathrm{K\,s^2\,m^{-1}})\,a,\nonumber\\
 a_\Delta&=(1.18353\times10^{10}\,\mathrm{m\,s^{-1}})\,\nu.
 \label{uc:numerical-conversion}
\end{align}
Table~\ref{uc:table} uses illustrative optical wavelengths and a $1\,\mathrm{GHz}$ mode. Electron and proton entries use free-particle gyromagnetic ratios from the 2022 CODATA adjustment; the cesium transition defines the SI second~\cite{NISTCODATA2022,NISTSI2019}.

\begin{table*}[t]
\caption{Reference gap and mode-energy scales, with $k_BT_U=\Delta E$ at $a_\Delta$; these are not carrier loss thresholds. Optical energies are not polarization splittings. Spin rows use a $1\,\mathrm T$ rest-frame field (unshielded proton). Values have six significant figures and no assigned decay rate.}
\label{uc:table}
\begin{ruledtabular}
\begin{tabular}{lccc}
Transition or selected mode
 & \(\nu\;(\mathrm{Hz})\)
 & \(\Delta E/k_B\;(\mathrm K)\)
 & \(a_\Delta\;(\mathrm{m\,s^{-2}})\)\\
\hline
Optical mode, \(800\,\mathrm{nm}\)
 & \(3.74741\times10^{14}\) & \(1.79847\times10^4\) & \(4.43518\times10^{24}\)\\
Optical mode, \(1550\,\mathrm{nm}\)
 & \(1.93414\times10^{14}\) & \(9.28243\times10^3\) & \(2.28912\times10^{24}\)\\
Microwave mode, \(1\,\mathrm{GHz}\)
 & \(1.00000\times10^9\) & \(4.79924\times10^{-2}\) & \(1.18353\times10^{19}\)\\
Free-electron Zeeman splitting, \(1\,\mathrm T\)
 & \(2.80250\times10^{10}\) & \(1.34499\) & \(3.31685\times10^{20}\)\\
Unshielded-proton Zeeman splitting, \(1\,\mathrm T\)
 & \(4.25775\times10^7\) & \(2.04340\times10^{-3}\) & \(5.03919\times10^{17}\)\\
\(^{133}\mathrm{Cs}\) ground-state hyperfine transition
 & \(9.19263\times10^9\) & \(4.41177\times10^{-1}\) & \(1.08798\times10^{20}\)
\end{tabular}
\end{ruledtabular}
\end{table*}

\subsection{Photonic encodings and accelerated apparatus}
\label{uc:photons}

A photon follows a null trajectory, with no rest-frame proper time or
proper acceleration. Here $a$ describes an accelerated source, detector,
or material apparatus with a specified gap and sampled modes. The optical
entries use $\nu=c_{\mathrm{light}}/\lambda$ in the specified mode frame;
moving-detector protocols must include Doppler shifts and mode selection.

Same-frequency polarization states have equal free-field energy, so
their qubit splitting is not $h\nu$. A polarization qubit differs from a
two-level absorption detector, whose population is $r$. A thermal bosonic
mode at $T_U$ instead has mean occupation
$\bar n=[\exp(a_\Delta/a)-1]^{-1}$.

A photonic Bell prediction requires joint detection probabilities, including attenuation, dark counts, filtering, mode mismatch, and unresolved rotations. Loss retains its erasure outcome or uses the setting-independent herald of Sec.~\ref{tr:detectors}. Rindler-mode entanglement uses a different subsystem partition; its single-mode approximation holds only for restricted states~\cite{Bruschi2010} and does not establish a universal carrier ranking.

\subsection{Massive carriers and the interaction spectrum}
\label{uc:massive}

An internal massive-carrier clock follows a timelike worldline, but the common $mc_{\mathrm{light}}^2\id$ contribution cancels from relative phases and Gibbs populations: the gap and matrix elements determine its internal response. Mass also affects trajectory forces, recoil, and environment.

For a spin-$\tfrac12$ Zeeman pair, $\nu=|\gamma|B/(2\pi)$~\cite{NISTCODATA2022}. Table~\ref{uc:table} shows that the heavier proton has the smaller thermal comparison scale in this encoding. Hyperfine and optical transitions give other scales; shielding, binding, level mixing, and acceleration fields can alter the actual gap. Reference energies alone specify neither an accelerator nor a mass-ordered robustness law.

An ideal uniformly accelerated electric-dipole two-level detector in the electromagnetic vacuum has population rate~\cite{Moustos2018,ZhuYuLu2006}
\begin{equation}
 \Gamma_{1,\mathrm{EM}}
 =\Gamma_{0,\mathrm{EM}}F(a)\coth x,\qquad
 F(a)=1+\left(\frac{a}{c_{\mathrm{light}}\Omega}\right)^2.
 \label{uc:em-rate}
\end{equation}
Here $\Gamma_{0,\mathrm{EM}}$ is its inertial spontaneous-emission rate, while $r$ remains Eq.~\eqref{ur:eq:occupation}.
The matched inertial electric-dipole detector in a thermal field at
$T_U$ has rate $\Gamma_{0,\mathrm{EM}}\coth x$ without $F(a)$.
Thus the stationary population agrees while the accelerated-vacuum
relaxation rate differs~\cite{ZhuYuLu2006,Moustos2018}. The dipole approximation, prescribed trajectory, and controlled weak-coupling regime are assumed; its secular transverse generator without added dephasing gives $\Gamma_{2,\mathrm{EM}}=\Gamma_{1,\mathrm{EM}}/2$. This does not assign the formula to a magnetic spin or photon polarization.

At $a=a_\Delta$, $F=1+4\pi^2\simeq40.48$: equal temperature and inertial rate therefore do not imply equal lifetimes.

\subsection{From energy scales to conditional resource thresholds}
\label{uc:thresholds}

For independent identical scalar channels, the Bell input $\ket{\Phi^+}$ and equal proper exposure $\Gamma_0\tau=0.1$, Sec.~\ref{ur:subsec:inversion} gives
\begin{equation}
 \frac{a_{B,2}}{a_\Delta}\simeq1.68366,\qquad
 \frac{a_{E,2}}{a_\Delta}\simeq4.40681.
 \label{uc:normalized-thresholds}
\end{equation}
These are model boundaries with optimized single-copy CHSH readout, not measured thresholds for every row in Table~\ref{uc:table}. Acceleration units require the gap and seconds require $\Gamma_0$; changing the gap generally changes that rate.

At fixed $a$, independent identical electric-dipole channels with no added dephasing give
\begin{equation}
 \tau_{X,2}^{\mathrm{EM}}(a)
 =\frac{\tau_{X,2}^{\mathrm{scalar}}(a;\Gamma_{0,\mathrm{EM}})}{F(a)},
 \qquad X\in\{B,E\},
 \label{uc:em-lifetimes}
\end{equation}
where the numerator is Eq.~\eqref{ur:eq:pair-lifetimes} evaluated with $\Gamma_{0,\mathrm{EM}}$. Recomputing the acceleration boundaries at $\Gamma_{0,\mathrm{EM}}\tau=0.1$ gives
\begin{align}
 a_{B,2}^{\mathrm{EM}}/a_\Delta&\simeq0.244102,\nonumber\\
 a_{E,2}^{\mathrm{EM}}/a_\Delta&\simeq0.464049.
 \label{uc:em-threshold-example}
\end{align}
Each lifetime is its scalar function times the positive increasing factor $x^2/(x^2+\pi^2)$, so the same endpoint argument ensures unique roots. These compare ideal models at fixed normalized exposure, not devices at a measured common lifetime.

With $\eta_{0,\mathrm{EM}}=\Gamma_{0,\mathrm{EM}}/\Omega$, secular validity requires
\begin{equation}
 \eta_{0,\mathrm{EM}}(1+4\pi^2\alpha^2)
 \coth[1/(2\alpha)]\ll1,
 \label{uc:em-secular-window}
\end{equation}
or $8\pi^2\eta_{0,\mathrm{EM}}\alpha^3\ll1$ at large $\alpha$. Taking $\eta_{0,\mathrm{EM}}=10^{-4}$ and $\Gamma_{0,\mathrm{EM}}\tau=0.1$ gives $\Omega\tau=10^3$ and, at the respective boundaries, $\Gamma_{1,\mathrm{EM}}/\Omega\simeq3.47\times10^{-4}$ and $1.20\times10^{-3}$. These scale choices are not microscopic error estimates: physical certification needs a channel-error bound smaller than the resource margin (Secs.~\ref{op:certification} and~\ref{ur:subsec:robustness}). The kinematic qualification of Sec.~\ref{ur:subsec:scope} also applies. Decimal precision describes numerical roots only.

These lifetimes require uniform linear acceleration, a resolved positive
gap, and the switching/readout conditions of Sec.~\ref{ur:subsec:scope}.
Circular trajectories, degenerate encodings, and freely propagating
uncoupled photons require different models.

\section{Discussion and conclusion}
\label{sec:conclusion}
For recorded phase noise on a depolarized Bell pair, the available
measurement output limits Bell certification. Some records preserve CHSH
violation when Bob retains them, yet every deterministic compression of
his qubit and record into one qubit loses it. The sharp metric-only
intervals quantify a separate limitation of an aggregate information
summary.

With unrestricted record size, that limitation persists at every finite
order of local geometry. The constructed readouts share a phase prior,
uniform flag alphabet, and diagonal references. Their pulled-back metrics
and contrasts agree to the prescribed orders, including curvature in the
explicit example, while their optimal compressed Bell outcomes differ.
Complete transverse entropy response, or sufficiently many moments under
a known finite support bound, determines the reliability law and both
CHSH optima. This identifies the missing channel information within the
model; the high-order construction has shrinking accuracy margins.

A circular Berry loop saturates the known isoholonomic action bound and
reduces the control budget at fixed ideal channel. Zero-expectation
conditional errors give a quadratic reduced-channel certificate after
ancillary discarding; a nonzero phase expectation can restore first-order
sensitivity. The implementation and recovery comparisons do not identify
a unique geometric origin from the final channel.

The transport and thermal applications distinguish Bell violation,
entanglement, and channel entanglement breaking. The finite-memory phase
model has a quartic onset and a controlled cubic limit. Accelerated-detector
predictions specialize known channels and require resource margins larger
than model error; a microscopic switching or finite-coupling budget remains
unresolved. Carrier comparisons require encoded gaps, interaction spectra,
clocks, and retained observations. These results fix neither a universal
separation lifetime nor a detector-independent acceleration bound.

\begin{samepage}
\section*{Data availability}
This theoretical work reports no experimental data. Analytical figure data and the generation and verification scripts accompany the manuscript source.
\end{samepage}

\appendix

\section{Geometric constructions and supplementary calculations}
\label{app:geometry}
\subsection{One Hessian metric and two affine connections}
\label{eg:subsec_hessian}

Let \(M\) have affine coordinates \(x^a\) and a smooth strictly convex
potential \(F\), with positive-definite Hessian. The Bregman contrast and
metric are
\begin{align}
 B_F(x,y)
 &=F(x)-F(y)-\partial_aF(y)(x^a-y^a), \label{eg:bregman}\\
 g_{ab}&=\partial_a\partial_bF. \label{eg:hessian_metric}
\end{align}
Repeated coordinate indices are summed. The affine connection
\(\nabla\) has vanishing coefficients in \(x\); its metric-dual connection
\(\nabla^*\) is defined by
\begin{equation}
 Xg(Y,Z)=g(\nabla_XY,Z)+g(Y,\nabla_X^*Z).
 \label{eg:duality}
\end{equation}
Both connections are flat and torsion free on a Hessian chart.
The coordinates \(\eta_a=\partial_aF\) are affine for \(\nabla^*\).
On the corresponding Legendre domain,
\begin{equation}
 g_\eta=\operatorname{Hess}F^*=g_x^{-1},
 \qquad \dd\eta=g_x\,\dd x.
 \label{eg:dual_components}
\end{equation}
Thus duality gives two coordinate descriptions of one metric
\cite{AmariNagaoka2000}. The connections differ through the cubic tensor:
\begin{align}
 C&=\nabla g,\qquad D=\tfrac12(\nabla+\nabla^*),\nonumber\\
 K&=\tfrac12(\nabla-\nabla^*)=-\tfrac12 C^\sharp.
 \label{eg:midpoint}
\end{align}
Here \(g(C_X^\sharp Y,Z)=C(X,Y,Z)\). The connection \(D\) is
Levi-Civita. With curvature convention
\(R^D(X,Y)=[D_X,D_Y]-D_{[X,Y]}\), dual flatness gives
\begin{equation}
 R^D(X,Y)=-[K_X,K_Y]
         =-\tfrac14[C_X^\sharp,C_Y^\sharp].
 \label{eg:cubic_curvature}
\end{equation}
For the sign, adding the vanishing curvatures of \(D\pm K\) cancels
the derivative terms and gives \(0=2R^D+2[K_X,K_Y]\).
Thus dual flatness allows a curved midpoint precisely when the cubic
endomorphisms do not commute.

\subsection{Gibbs coordinates and contrast orientation}
For $F(\rho)=\Tr\rho\log\rho$, Eq.~\eqref{eg:bregman} gives
$\DU(\rho\Vert\sigma)=B_F(\rho,\sigma)$ on the faithful manifold.
Mixture coordinates use the affine trace-one hyperplane. An alternative
global chart uses trace-free Hermitian \(X\):
\begin{equation}
 \rho_X=\frac{e^{-X}}{\Tr e^{-X}},\qquad
 \Psi(X)=\log\Tr e^{-X}.
 \label{eg:gibbs_chart}
\end{equation}
The inverse is \(X=-\log\rho+(\Tr\log\rho)\id/n\). In this chart the
pullback of \(\gBKM\) is \(\operatorname{Hess}\Psi\), while the contrast
has the reversed Bregman orientation
\begin{equation}
 \DU(\rho_X\Vert\rho_Y)=B_\Psi(Y,X).
 \label{eg:gibbs_orientation}
\end{equation}
Indeed, \(\dd\Psi_X(H)=-\Tr(\rho_XH)\); substituting
\(\log\rho_X=-X-\Psi(X)\id\) proves the orientation, including for
noncommuting faithful states.

\subsection{Phase completion of a probability simplex}
\label{eg:sec_simplex}

For the simplex interior $\Delta_{n-1}^{\circ}$, doubling the $n-1$
tangent coordinates matches the real dimension of $\mathbb{CP}^{n-1}$.
Use exponential coordinates with potential
\begin{align}
 p_i&=\frac{e^{\theta_i}}{Z}\quad(i<n),&
 p_n&=\frac1Z,\\
 Z&=1+\sum_{i<n}e^{\theta_i},&
 \psi(\theta)&=\log Z.
 \label{eg:simplex_coordinates}
\end{align}
The Fisher matrix is
\(g_{ij}=\psi_{ij}=p_i\delta_{ij}-p_ip_j\), where
\(i,j<n\).
Let \(u_i\) denote tangent-fiber coordinates in this exponential
affine chart. The Dombrowski construction equips the tangent bundle
with
\begin{align}
 h_B&=\psi_{ij}
       (\dd\theta_i\otimes\dd\theta_j+\dd u_i\otimes\dd u_j),\\
 I_B\partial_{\theta_i}&=\partial_{u_i},&
 I_B\partial_{u_i}&=-\partial_{\theta_i},\\
 \omega_B&=\psi_{ij}\,\dd\theta_i\wedge\dd u_j.
 \label{eg:dombrowski}
\end{align}
Symmetry of \(\psi_{ijk}\) gives \(\dd\omega_B=0\), and the displayed
complex structure is integrable. These Kähler data admit the following
adaptation of Molitor's projective realization.

\begin{proposition}[Local projective realization]
\label{eg:prop_simplex}
With the convention \(\gFS=\operatorname{Re}Q\) for the pure-state
quantum geometric tensor, the map
\begin{equation}
 \mathcal T(\theta,u)=
 \left[e^{(\theta_1+\ii u_1)/2}:\cdots:
       e^{(\theta_{n-1}+\ii u_{n-1})/2}:1\right]
 \label{eg:simplex_map}
\end{equation}
is holomorphic and satisfies
\begin{equation}
 \mathcal T^*\gFS=\tfrac14 h_B,\qquad
 \mathcal T^*\omega_{\mathrm{FS}}=\tfrac14\omega_B,
 \label{eg:simplex_pullback}
\end{equation}
where \(\omega_{\mathrm{FS}}(X,Y)=\gFS(IX,Y)\).
\end{proposition}

This is Molitor's realization in exponential coordinates
\cite[Lemma 3.1 and Proposition 3.3]{Molitor2012}.
Writing \(\zeta_i=(\theta_i+\ii u_i)/2\) gives
\(\dd\zeta_i=(\dd\theta_i+\ii\dd u_i)/2\), accounting for the
factor \(1/4\), which Molitor instead includes in his Fisher metric.

This established realization uses a classical simplex. Doubling the full
faithful matrix manifold instead gives complex dimension $n^2-1$, not
the original ray space.

\subsubsection{The phase lattice, completion, and line connection}
\label{eg:subsec_lattice}

Equation~\eqref{eg:simplex_map} is a covering map rather than a global
one-to-one correspondence. Its deck transformations are
\begin{equation}
 u_i\longmapsto u_i+4\pi k_i,\qquad k_i\in\mathbb Z.
 \label{eg:phase_lattice}
\end{equation}
The quotient is the open dense locus \((\mathbb C^*)^{n-1}\) where all
homogeneous coordinates are nonzero \cite[Lemma 3.1]{Molitor2012}.
Full projective space additionally adjoins zero-amplitude strata,
identifying phases whose amplitudes vanish. These global specifications
go beyond the local Hessian equations.

A normalized section on the displayed chart is
\begin{equation}
 \ket{\psi(\theta,u)}
 =\sum_{i<n}\sqrt{p_i}\,e^{\ii u_i/2}\ket{i}
   +\sqrt{p_n}\ket{n}.
 \label{eg:simplex_section}
\end{equation}
It yields the real Berry connection and curvature
\begin{align}
 A&=\ii\langle\psi|\dd\psi\rangle
   =-\tfrac12\sum_{i<n}p_i\,\dd u_i,\\
 \mathcal F&=\dd A
   =-\tfrac12\psi_{ij}\,\dd\theta_i\wedge\dd u_j
   =-2\mathcal T^*\omega_{\mathrm{FS}}.
 \label{eg:simplex_berry}
\end{align}
The Hessian thus controls both tensors after the phase variables and
ray interpretation have been supplied. A gauge change
\(\ket\psi\mapsto e^{\ii\chi}\ket\psi\) gives \(A\mapsto A-\dd\chi\)
and leaves \(\mathcal F\) unchanged. For a loop bounding a surface
in one trivialization, \(\gamma=\oint A=\int\mathcal F\pmod{2\pi}\);
multiple patches require
transition functions. On completed projective space the line bundle
can be nontrivial, with integrally normalized first-Chern curvature
periods. These dimensionless data fix no action unit. An adiabatic
phase experiment also requires a Hamiltonian and transport protocol
\cite{Berry1984,Simon1983}.

\subsection{Projector geometry and quantum transport}
\label{eg:sec_projector}

More generally, let \(P(\lambda)\) be a smooth constant-rank family of
orthogonal projectors on the supplied Hilbert space. Its ranges form
a vector bundle over the parameter manifold \(\mathcal B\).
A spectral eigenvalue or cluster defining \(P\) must remain isolated.

Write \(Q=\id-P\). The projected connection and the horizontal
projector derivative are
\begin{equation}
 \nabla^P=P\,\dd,\qquad
 \mathcal U_X=Q(\dd P)_XP.
 \label{eg:projector_connection}
\end{equation}
Their operator-valued geometric tensor is
\begin{equation}
 \mathcal G(X,Y)
 =\mathcal U_X^\dagger\mathcal U_Y
 =P(\dd P)_XQ(\dd P)_YP.
 \label{eg:operator_qgt}
\end{equation}
Differentiating \(P^2=P\) gives \(P(\dd P)P=0\). Consequently
\begin{align}
 R^P(X,Y)
 &=P[(\dd P)_X,(\dd P)_Y]P\\
 &=\mathcal G(X,Y)-\mathcal G(Y,X), \label{eg:projector_curvature}\\
 g^P(X,Y)
 &=\operatorname{Re}\Tr\mathcal G(X,Y)
 =\tfrac12\Tr\!\left[(\dd P)_X(\dd P)_Y\right].
 \label{eg:projector_metric}
\end{align}
Equation~\eqref{eg:projector_metric} is positive semidefinite on
parameter space and is a metric when the projector map is an
immersion. These formulas express metric response and connection
curvature through different parts of one tensor
\cite{ProvostVallee1980,GauvinCompletion2026}.

\subsubsection{Area bound and degenerate sectors}
\label{eg:subsec_rank}

For rank one, write $P=\ket\psi\bra\psi$ and use the conventions
in Eqs.~\eqref{eg:ray_qgt} and~\eqref{eg:ray_curvature}.
For real tangents \(X,Y\), positivity of the Gram matrix of
\((\id-P)\partial_X\ket\psi\) and \((\id-P)\partial_Y\ket\psi\) gives
\begin{equation}
 |\mathcal F(X,Y)|
 \leq 2\sqrt{\gFS(X,X)\gFS(Y,Y)-\gFS(X,Y)^2}.
 \label{eg:qgt_area_bound}
\end{equation}
The factor two follows from Eq.~\eqref{eg:ray_curvature}; the area
is that of the same ray family's Fubini-Study metric. A bound using
another statistical metric requires a separate comparison.

For rank \(r>1\), choose a local orthonormal frame
\(\{\ket{v_a}\}_{a=1}^r\). The anti-Hermitian connection matrix,
curvature, and parallel transport are
\begin{align}
 a_{ab}&=\langle v_a|\dd v_b\rangle,&
 f&=\dd a+a\wedge a,\\
 U_\gamma&=\mathcal P\exp\!\left(-\int_\gamma a\right).
 \label{eg:wilczek_zee}
\end{align}
A frame change \(v\mapsto vW\), \(W\in U(r)\), gives
\(a\mapsto W^\dagger aW+W^\dagger\dd W\).
Loop holonomy changes by conjugation and generally requires path
ordering \cite{WilczekZee1984}; a scalar phase needs a line sector or
Abelian reduction. An adiabatic implementation additionally specifies
a gapped Hamiltonian, preparation and readout in the fiber, and treatment
of dynamical phases.

\subsection{Spectral response as a comparison of the two geometries}
\label{eg:subsec_spectral_response}

For a faithful family with simple spectrum on a parameter patch, write
\begin{equation}
 \rho(\lambda)=\sum_{a=1}^n p_a(\lambda)P_a(\lambda),
 \qquad P_a=\ket{a}\bra{a},
 \label{eg:spectral_model}
\end{equation}
and choose a smooth local orthonormal eigenframe. Put
\(\alpha_{ab}(X)=\langle a|\partial_X b\rangle\).
Evaluation of the kernel in Eq.~\eqref{eg:bkm_kernel} gives
\begin{align}
 \gBKM_\rho(\partial_X\rho,\partial_Y\rho)
 &=\sum_a\frac{(Xp_a)(Yp_a)}{p_a}
 \nonumber\\
 &\quad+2\sum_{a<b}(p_a-p_b)(\log p_a-\log p_b)
 \nonumber\\[-2pt]
 &\hspace{25pt}\times
 \operatorname{Re}\!\left[
 \overline{\alpha_{ab}(X)}\,\alpha_{ab}(Y)\right].
 \label{eg:bkm_spectral_split}
\end{align}
The diagonal derivative is \(Xp_a\), while
\((\partial_X\rho)_{ab}=(p_b-p_a)\alpha_{ab}(X)\) for \(a\ne b\).
Substituting in Eq.~\eqref{eg:bkm_kernel} and pairing conjugate entries
gives the nonnegative weights and factor two. The first term is the
Fisher response of the eigenvalues; the second weights projector motion
by their populations. The corresponding eigenline tensor is
\begin{equation}
 Q^{(a)}(X,Y)
 =\sum_{b\ne a}
   \overline{\alpha_{ba}(X)}\,\alpha_{ba}(Y).
 \label{eg:eigenprojector_tensor}
\end{equation}
BKM weights the real parts of these eigenframe derivatives by the
spectrum; eigenline Berry curvature uses their imaginary parts. Only
projector motion remains on a fixed-spectrum orbit. Individual eigenlines
require simple spectrum; at positive degeneracies BKM stays smooth,
while an isolated cluster uses Eq.~\eqref{eg:wilczek_zee}. Boundary limits
retain the support qualifications of Sec.~\ref{eg:subsec_umegaki}.

\subsection{Coordinate and gauge checks for the qubit family}
\label{app:qubit-coordinate-details}
For Eq.~\eqref{qm:eq:state}, direct differentiation in coordinate order
$(\vartheta,\phi)$ gives the quantum geometric tensor
\begin{equation}
 Q=\begin{pmatrix}
 1&\ii\sin\vartheta\cos\vartheta\\
 -\ii\sin\vartheta\cos\vartheta&\sin^2\vartheta\cos^2\vartheta
 \end{pmatrix},
 \label{qm:eq:qgt}
\end{equation}

In probability coordinates the same formulas read
\begin{align}
 \dd s_{FS}^2&=\frac{\dd p^2}{4p(1-p)}
                 +p(1-p)\,\dd\phi^2,\nonumber\\
 A&=-p\,\dd\phi,\qquad \mathcal F=-\dd p\wedge\dd\phi.
 \label{qm:eq:prob-phase}
\end{align}
The first term is one quarter of the Bernoulli Fisher metric; the remaining terms require the supplied relative phase.

The section in Eq.~\eqref{qm:eq:state} is regular at $\vartheta=0$. Near the other pole use $\ket{\psi_S}=e^{-\ii\phi}\ket\psi$, giving
\begin{equation}
 A_S=A+\dd\phi=(1-p)\dd\phi.
 \label{qm:eq:south-section}
\end{equation}
Both sections give the same curvature and cyclic holonomy. Over the complete projective line, $\int\mathcal F=-2\pi$ in the stated orientation. With $\chi=2\vartheta$, the metric is one quarter of the unit-sphere metric and has Gaussian curvature $4$, while entanglement varies; see Sec.~\ref{sec:limits}.

\subsection{The fixed-spectrum metric factors}
\label{app:metric-comparison-proof}

\begin{proof}
At fixed $p_{\mathrm{mix}}$, $\dd\rho_{p_{\mathrm{mix}}}=v\dd P$ has only $a$-$b$
blocks. Equation~\eqref{eg:projector_metric} therefore reduces the
spectral metric to $2v^2c(a,b)\gFS$.
The kernels $c_{BKM}(a,b)=\ln(a/b)/(a-b)$ and
$c_{SLD}(a,b)=2/(a+b)$ give the stated factors; both have continuous
diagonal value $1/a$~\cite{PetzSudar1996}.
\end{proof}

\subsection{Thermodynamic calibration}
Equation~\eqref{qm:eq:parent} supplies an optional thermodynamic calibration. At inverse temperature $\beta>0$, its Gibbs state has
\begin{equation}
 a=\frac1{1+3e^{-\beta\Delta_E}},\qquad
 b=\frac{e^{-\beta\Delta_E}}{1+3e^{-\beta\Delta_E}},
 \label{qm:eq:gibbs-spectrum}
\end{equation}
so $\ln(a/b)=\beta\Delta_E$. For a fixed Gibbs reference, a unitary displacement obeys the exact identity
\begin{align}
 &\beta^{-1}\DU(U\rho_\beta U^\dagger\|\rho_\beta)\nonumber\\
 &\qquad=\Tr[H(U\rho_\beta U^\dagger-\rho_\beta)].
 \label{qm:eq:gibbs-calibration}
\end{align}
This follows from unitary entropy preservation and $\ln\rho_\beta=-\beta H-\ln Z$, with the supplied $H$ and $\beta$ fixing the energy calibration.

\section{Channel geometry and conditional information}
\label{app:channel-details}
\subsection{Output support and the loss Hessian}
Let $\Phi$ be a fixed CPTP map and evaluate output logarithms on
$\mathcal K=\operatorname{ran}\Phi(\id)$. For
$0<a\id\le\rho\le b\id$, positivity gives
\begin{equation}
 a\Phi(\id)\le\Phi(\rho)\le b\Phi(\id).
 \label{ch:support}
\end{equation}
Thus every faithful input has a faithful output on the fixed support
$\mathcal K$, where output entropy can be differentiated.

On the affine trace-one input space write
\begin{equation}
 F_\Phi(\rho)=\Tr\rho\ln\rho
              -\Tr(\Phi\rho)\ln(\Phi\rho).
 \label{ch:loss-potential}
\end{equation}
Linearity and trace preservation give $\mathcal L_\Phi=B_{F_\Phi}$, whose diagonal Hessian is Eq.~\eqref{ch:loss-metric}.

For a smooth faithful curve $\rho_s$ with $\rho_0=\rho$ and trace-zero
Hermitian tangent $\dot\rho_0=U$,
\begin{equation}
 \mathcal L_\Phi(\rho_s,\rho)
 =\tfrac12s^2G_{\Phi,\rho}(U,U)+O(s^3).
 \label{ch:loss-jet}
\end{equation}
The vanishing first differential removes dependence on curve acceleration. The output metric of Eq.~\eqref{ch:measurement} is Fisher information on active outcomes; both tensors in Eq.~\eqref{ch:loss-metric} are pulled to the input carrier.

\subsection{Composition, recovery, and a moving apparatus}
For a second fixed CPTP channel $\Psi$,
\begin{align}
 \mathcal L_{\Psi\circ\Phi}(\rho,\sigma)
 &=\mathcal L_\Phi(\rho,\sigma)
   +\mathcal L_\Psi(\Phi\rho,\Phi\sigma),
 \label{ch:cocycle}\\
 G_{\Psi\circ\Phi,\rho}(U,U)
 &=G_{\Phi,\rho}(U,U)
   +G_{\Psi,\Phi\rho}(\Phi U,\Phi U).
 \label{ch:metric-cocycle}
\end{align}
The defining difference telescopes, and differentiation gives the metric identity; data processing makes each loss nonnegative.

If one CPTP map $\mathcal R$ recovers every state of an embedded model, $\mathcal R\Phi\rho=\rho$, data processing in both directions equates the restricted contrasts and hence their metric and connection data on regular domains~\cite{Petz1986,GauvinCompletion2026}. Injectivity alone does not guarantee such recovery.

These results hold the channel fixed. A moving apparatus instead gives
\begin{equation}
 \dd(\Phi_\lambda\rho_\lambda)
 =\Phi_\lambda(\dd\rho_\lambda)
  +(\dd\Phi_\lambda)\rho_\lambda.
 \label{ch:moving}
\end{equation}
The apparatus term requires a separate comparison.

\subsection{A variance identity for branch-dependent response}
Forgetting the label of a fixed ensemble of unitary transports gives the channel
\begin{equation}
 \Lambda(X)=\sum_j w_j U_j X U_j^\dagger,
 \qquad w_j\ge0,\quad \sum_j w_j=1.
 \label{op:mixture}
\end{equation}
Assume each branch preserves the same faithful reference state,
\begin{equation}
 U_j\rho U_j^\dagger=\rho.
 \label{op:preserve}
\end{equation}

\begin{proposition}[Loss as branch-response variance]
Under Eqs.~\eqref{op:mixture} and~\eqref{op:preserve}, every trace-zero Hermitian tangent $X$ satisfies
\begin{align}
 G_{\Lambda,\rho}(X,X)
  &=\frac12\sum_{j,k}w_jw_k\|Y_j-Y_k\|_\rho^2,
 \label{op:variance}\\
 Y_j&=U_j X U_j^\dagger,\qquad
 \|Y\|_\rho^2=\gBKM_\rho(Y,Y).
 \nonumber
\end{align}
\end{proposition}
\begin{proof}
Unitary invariance of the BKM metric and Eq.~\eqref{op:preserve} give $\|Y_j\|_\rho^2=\|X\|_\rho^2$. The channel fixes $\rho$, so Eq.~\eqref{ch:loss-metric} becomes
\begin{equation}
 G_{\Lambda,\rho}(X,X)
 =\sum_jw_j\|Y_j\|_\rho^2
  -\left\|\sum_jw_jY_j\right\|_\rho^2.
 \label{op:variance-proof}
\end{equation}
Expanding the squared norms proves Eq.~\eqref{op:variance}.
\end{proof}
This standard inner-product variance identity holds for any metric preserved by the branches at a common reference; BKM supplies the channel-loss interpretation.

For the phase channel in Eq.~\eqref{op:phase-channel}, each unitary
branch preserves any diagonal faithful qubit reference.
A transverse Hermitian tangent $X=x\sigma_x+y\sigma_y$ obeys $\Lambda X=\cos\delta\,X$, whereas diagonal tangents are unchanged. Therefore
\begin{equation}
 G_{\Lambda,\rho}(X,X)
 =\sin^2\delta\,\gBKM_\rho(X,X).
 \label{op:phase-loss}
\end{equation}
At $\delta=\pi/2$ the transverse response is erased; at $\delta=\pi$ the conjugation branches agree, giving a deterministic phase flip and zero loss.

\subsection{Proof and coarse-graining of the record bounds}
\label{rec:proof}
Write $\sigma=\operatorname{diag}(s,1-s)$, $0<s<1$. In its eigenbasis the
BKM kernel is $c(a,b)=(\ln a-\ln b)/(a-b)$, with $c(a,a)=1/a$.
If $X_{01}=z$, then
\begin{equation}
 \gBKM_\sigma(X,X)=2c(s,1-s)|z|^2.
 \label{rec:kernel}
\end{equation}
The $y$ block has reference $p_y\sigma$ and tangent
$p_y\Phi_y(X)$, whose off-diagonal magnitude is $p_yq_y|z|$.
Since $c(p_ya,p_yb)=p_y^{-1}c(a,b)$, its contribution is
$p_yq_y^2\gBKM_\sigma(X,X)$. Summing gives
Eq.~\eqref{rec:metric}. The same homogeneity holds for a normalized
monotone-metric kernel; no singular reference or pure-state BKM limit is
used. A diagonal feedback rotates $\mu_y$ without changing its magnitude,
and
$|\sum_yp_ye^{\ii\alpha_y}\mu_y|\le\sum_yp_y|\mu_y|$,
with equality when all nonzero terms align. Hence its optimal unflagged
multiplier is $m$. The inequalities $q_y^2\le q_y$ and
$(\sum p_yq_y)^2\le\sum p_yq_y^2$ prove Eq.~\eqref{rec:bounds}.
The lower equality requires $q_y\in\{0,1\}$ almost surely; the upper
requires constant $q_y$ almost surely. The erasure and binary examples in
the main text realize these cases at every $0\le\eta\le1$ with the same
uniform phase prior.

Classically degrading the record by a fixed stochastic map $Y\to Z$
gives $\mu_z=\mathbb E(\mu_Y|Z=z)$, by the Markov assumption
$J\to Y\to Z$. Conditional variance and the triangle inequality yield
\begin{align}
 \eta_Y-\eta_Z
 &=\mathbb E|\mu_Y-\mathbb E(\mu_Y|Z)|^2\ge0,\nonumber\\
 m_Z&\le m_Y.
 \label{rec:garbling}
\end{align}
Equality in the first line holds exactly when $\mu_Y$ is almost surely
determined by $Z$. Equality in the second holds exactly when, within
each nonzero-probability $Z$ block, the nonzero $\mu_y$ have a common
complex argument. Thus neither metric retention nor optimally corrected
coherence can improve by degrading a fixed record. This comparison is made
before feedback; correcting with a finer record and subsequently deleting
that record is a different physical operation.

The corrected $\rho_v$ is an $X$ state with middle diagonal entries
$(1-v)/4$, corner coherence $v\mathcal C_0m/2$, and correlation singular
values $v,v\mathcal C_0m,v\mathcal C_0m$. The concurrence, PPT, and
Horodecki criteria give Eq.~\eqref{rec:resources}; monotonicity in $m$
gives the intervals. The two endpoint records prove their optimality
over this phase-record class for fixed $\eta,v,\mathcal C_0$.
For $0\le m_1\le m_2$ and $m_2>0$, the local incoherent channel
$\Lambda_r=(1+r)\mathrm{id}/2+(1-r)\operatorname{Ad}_{Z}/2$,
$r=m_1/m_2$, maps $\rho_{v,m_2}$ to $\rho_{v,m_1}$.
Coherence monotonicity gives Eq.~\eqref{rec:coherence-bounds}; when
$m_2=0$, both states coincide. The same records attain the endpoints.
These ideal-channel bounds include no coherence continuity estimate,
and the endpoints need not be accessible for a separately fixed prior
or restricted readout apparatus.

\subsection{Proof of the record and compression optima}
\label{rec:compression-proof}
After alignment the flagged state has blocks
\begin{equation}
 \rho_{v,q}=\tfrac14[\id_4+v(q\sigma_x\otimes\sigma_x
 -q\sigma_y\otimes\sigma_y+\sigma_z\otimes\sigma_z)].
 \label{rec:aligned-block}
\end{equation}
Each block has marginals $\id_2/2$ and correlation singular values
$v,vq,vq$. Only diagonal flag blocks of Bob's observables contribute.
For fixed Alice spin observables, the paired Bob operators are traceless,
so their binary-measurement extrema can be attained by spin observables.
The Horodecki criterion bounds each block by $2v\sqrt{1+q_y^2}$;
the common Alice and conditional Bob axes stated after
Proposition~\ref{rec:prop:compression} attain all these bounds.
For arbitrary binary POVMs it suffices to consider extreme Hermitian
contractions on Alice: spin observables or $\pm\id_2$. If either is
deterministic, Alice's observables commute and CHSH is at most $2$,
which deterministic outputs attain. This proves the retained-record
formula, with the maximum taken after averaging because Alice has no flag.

Any CPTP map on the block-diagonal input restricts to qubit CPTP maps
$\mathcal R_y(X)=\mathcal R(X\otimes\ket y\bra y)$.
Write their affine Bloch form as $r\mapsto M_yr+t_y$.
Trace-distance contraction for antipodal states implies
$\|M_y\|_{\rm op}\le1$. The translations contribute no correlations
because the conditional Alice marginals are $\id_2/2$. Thus
\begin{equation}
 T_{\rm out}=\sum_yp_yT_yM_y^{\mathsf T}.
\end{equation}
Since $M_y^{\mathsf T}M_y\preceq\id_3$,
\[
 (T_yM_y^{\mathsf T})(T_yM_y^{\mathsf T})^{\mathsf T}
 =T_yM_y^{\mathsf T}M_yT_y^{\mathsf T}\preceq T_yT_y^{\mathsf T}.
\]
The first two singular values of each summand before weighting are
therefore at most $v,vq_y$. Triangle inequalities
for the operator norm and the sum of the two largest singular values give
\begin{equation}
 s_1(T_{\rm out})\le v,\qquad
 s_1(T_{\rm out})+s_2(T_{\rm out})\le v(1+m).
 \label{rec:compression-singular}
\end{equation}
For $0\le s_2\le s_1$, these constraints imply
$s_1^2+s_2^2\le v^2(1+m^2)$: the extreme points are
$(0,0)$, $(v,0)$, $(v,vm)$, and
$(v(1+m)/2,v(1+m)/2)$, and the last has no larger squared norm than
$(v,vm)$. Phase alignment and deletion attain
$T_{\rm out}=v\operatorname{diag}(m,-m,1)$.
The two-qubit spin criterion and the deterministic binary value $2$
therefore prove the compression formula. The argument makes no
concurrence-optimality claim for arbitrary CPTP recovery.

If $\|\Omega^{\rm act}_{ABY}-\Omega_{ABY}\|_1\le\epsilon$, every fixed
retained CHSH expectation changes by at most $2\sqrt2\epsilon$.
Trace-norm contraction gives the same bound after each compression,
uniformly over maps and subsequent spin measurements. This proves both
examples' error margins without assuming maximally mixed marginals for
the perturbed state.

\subsection{Proof of the finite-order geometric obstruction}
\label{rec:finite-proof}
Let $g^f$ be any normalized Petz monotone metric~\cite{PetzSudar1996}.
Direct-sum homogeneity and fixed output-unitary invariance give the
pullbacks
\begin{align}
 G^f_\rho(X,Y)
 &=\int g^f_{\Lambda_q\rho}(\Lambda_qX,\Lambda_qY)\,\dd\nu(q),\nonumber\\
 D_\nu(\rho,\tau)
 &=\int\DU(\Lambda_q\rho\Vert\Lambda_q\tau)\,\dd\nu(q).
 \label{rec:pullback-law}
\end{align}
At a diagonal reference, $r$ coordinate derivatives of $G^f$ are
multilinear in $r$ base variations and two tangent slots. Each transverse
slot supplies one factor of $q$, while a diagonal slot is unchanged.
Simultaneous conjugation by $\sigma_z$ fixes the reference and reverses
all transverse slots; unitary invariance therefore removes odd powers.
The derivative depends only on
$M_0,\ldots,M_{\lfloor(r+2)/2\rfloor}$.
At a faithful diagonal pair the same argument makes each total contrast
derivative of order $k$ depend only on
$M_0,\ldots,M_{\lfloor k/2\rfloor}$.
The kernels are smooth on faithful states, including the maximally mixed
state. Equal $M_1,\ldots,M_N$ give the claimed jet equalities. Subtracting
from the common input contrast gives the same statement for its loss.

Set $L=N+1$ and $q_i=\sqrt{i/L}$ for $i=0,\ldots,L$.
The parity laws
\begin{align}
 \nu_-(q_i)&=2^{-N}\binom Li\quad(i\text{ even}),\nonumber\\
 \nu_+(q_i)&=2^{-N}\binom Li\quad(i\text{ odd})
 \label{rec:parity-laws}
\end{align}
are normalized and have equal $M_0,\ldots,M_N$ by
$\sum_{i=0}^L(-1)^i\binom Li i^k=0$ for $k<L$.
Both have $M_1=1/2$. Writing $M_k^\pm=\mathbb E_{\nu_\pm}q^{2k}$,
the first unmatched moment is
\[
 M_{N+1}^+-M_{N+1}^-
 =\frac{(-1)^N(N+1)!}{2^N(N+1)^{N+1}}\ne0.
\]
The integral representation
$\sqrt u=(2\sqrt\pi)^{-1}\int_0^\infty(1-e^{-ut})t^{-3/2}\dd t$
gives
\begin{equation}
 \delta_N:=m_+-m_-=
 \frac{\displaystyle\int_0^\infty(1-e^{-s})^Ls^{-3/2}\dd s}
 {2^{N+1}\sqrt{\pi L}}>0.
 \label{rec:jet-gap}
\end{equation}
The integral converges at both endpoints.
Any finite law with masses $w_i$ is realized by the same uniform prior
$J=\pm1$, phases $J\pi/2$, and readout
\begin{equation}
 p(i,b|j)=w_i(1+bj q_i)/2,\qquad b=\pm1.
 \label{rec:parity-readout}
\end{equation}
Its outcome probability is $w_i/2$ and posterior phase moment is
$\ii bq_i$. Split each weight $2^{-N}\binom Li$ into $\binom Li$
equal-weight copies. Each parity then has $2^N$ reliability labels and,
after pairing $b=\pm1$, the same uniform alphabet of $2^{N+1}$ flags.
A fixed relabeling identifies the output carriers and their diagonal
reference states exactly. All probabilities are independent of the
quantum input, and every trial is retained.

Choose $(1+m_+^2)^{-1/2}<v<(1+m_-^2)^{-1/2}$ and apply
Proposition~\ref{rec:prop:compression}. Its spin optima straddle $2$,
proving Proposition~\ref{rec:prop:finite-jets}. The construction has no
uniform margin: bounding $(1-e^{-s})^L\le\min(s^L,1)$ in
Eq.~\eqref{rec:jet-gap} gives
\begin{equation}
 0<\delta_N\le
 \frac{2+(L-1/2)^{-1}}{2^{N+1}\sqrt{\pi L}}.
\end{equation}
Since $|\partial_m[2v\sqrt{1+m^2}]|\le\sqrt2$ on this domain,
the displayed Bell separation also tends to zero as $N$ increases.

To verify the exact orders and the complete-response statement, the
commuting family of Eq.~\eqref{rec:complete-response} has
\begin{align}
 H_\nu''(t)
 &=\int\frac{q^2}{1-q^2t^2}\,\dd\nu(q)\nonumber\\
 &=\sum_{k\ge1}M_kt^{2k-2}.
 \label{rec:radial-metric}
\end{align}
Every normalized monotone metric agrees with this radial Fisher metric.
Integrating twice with $H_\nu(0)=H_\nu'(0)=0$ proves
Eq.~\eqref{rec:complete-response}. The parity laws first differ in
$M_{N+1}$, so the first possible metric and contrast discrepancies,
at orders $2N$ and $2N+2$, are attained.
Polynomial density on $[0,1]$ makes the full moment sequence determine
the law of $q^2$, hence $\nu$ because $q\mapsto q^2$ is one-to-one there.
For two laws with at most $K$ atoms each, their difference has at most
$2K$ support points; interpolation polynomials of degree at most $2K-1$
isolate every weight. This proves the finite-support qualification.
Neither result identifies a unique microscopic phase prior or implementation.

\subsection{Instruments, conditional states, and no signalling}
A POVM specifies outcome probabilities; an instrument also specifies
the state update. Let $\mathcal I_{a|x}$ be completely positive maps whose sum is trace preserving. Then
\begin{equation}
 p(a|x)=\Tr\mathcal I_{a|x}(\rho),\qquad
 \rho_{a|x}=\frac{\mathcal I_{a|x}(\rho)}{p(a|x)}
 \label{ch:instrument}
\end{equation}
when $p(a|x)>0$. Keeping the outcome gives a classical-quantum record; discarding it gives the nonselective channel $\sum_a\mathcal I_{a|x}$.

For Alice's local instrument on $\rho_{AB}$, Bob's unnormalized conditional states are
\begin{equation}
 \sigma_{a|x}=\Tr_A[(\mathcal I_{a|x}\otimes\mathrm{id})(\rho_{AB})].
 \label{ch:assemblage}
\end{equation}
Kraus completeness under the partial trace gives
\begin{equation}
 \sum_a\sigma_{a|x}=\Tr_A\rho_{AB}=\rho_B.
 \label{ch:nosignalling}
\end{equation}
Alice's choice leaves Bob's unconditional probabilities unchanged;
conditioning requires the outcome record.

Steering asks whether the assemblage has a local-hidden-state representation~\cite{WisemanJonesDoherty2007},
\begin{equation}
 \sigma_{a|x}=\sum_\lambda w_\lambda
                 p(a|x,\lambda)\tau_\lambda,
 \label{ch:lhs}
\end{equation}
with one ensemble of Bob states independent of $x$. An integral may replace the sum. For $\ket{\Phi^+}=(\ket{00}+\ket{11})/\sqrt2$, projective $Z$ and $X$ measurements give
\begin{align}
 \sigma_{0|Z}&=\tfrac12\ket0\bra0,&
 \sigma_{1|Z}&=\tfrac12\ket1\bra1,
 \nonumber\\
 \sigma_{\pm|X}&=\tfrac12\ket{\pm}\bra{\pm}.
 \label{ch:steering-example}
\end{align}
Positivity forces hidden states for the $Z$ assemblage onto its computational rays. Their diagonal mixtures cannot reproduce the $X$ rays: Eq.~\eqref{ch:lhs} fails although Eq.~\eqref{ch:nosignalling} holds.

\subsection{Proof of the reduced-channel error certificate}
\label{op:quadratic-proof}
For a fixed orientation history let $U(t)$ be the ideal ancillary
propagator and $\widetilde U(t)$ the actual one. Set
$W=U^\dagger\widetilde U$ and
$K=U^\dagger eU/\hbar$, so $\dot W=-\ii KW$.
The ideal state $U(t)\ket0$ follows the transported ray; hence
$K_{00}=0$. Write $W(t)\ket0=a(t)\ket0+\ket{b(t)}$, with
$\langle0|b(t)\rangle=0$, and define
$\ell(t)=\int_0^t\|K(s)\|\dd s$.
Duhamel's identity and unitarity imply
$\|b(t)\|\le\ell(t)$, while
$\dot a=-\ii\langle0|K|b\rangle$ gives
\begin{equation}
 |a(T)-1|\le\int_0^T\|K(t)\|\ell(t)\dd t
             =\ell(T)^2/2.
 \label{op:quadratic-amplitude}
\end{equation}
No assumption about the complementary diagonal block of $K$ is needed.

At a cyclic endpoint, ancillary tracing changes only the logical
off-diagonal block, by the ideal phase times $a(T)^*-1$. The resulting
channel difference has diamond norm $|a(T)-1|$. To verify the upper bound with any reference, the
off-diagonal part of a state $\rho$ is
$[\rho-(Z_B\otimes\id)\rho(Z_B\otimes\id)]/2$, whose trace norm is at
most one. Multiplication by a complex scalar changes its norm by the
scalar's modulus, up to a diagonal unitary. The input $\ket+_B$ attains
this value. Equation~\eqref{op:quadratic-amplitude}, the maximal channel
distance two, and $\ell(T)\le N\eta_H$ prove
Eq.~\eqref{op:cd-quadratic-error}. Averaging fixed branch histories or
applying fixed classical readout to their orthogonal flags preserves
the bound by convexity and contraction.

With an ideal reset to $\ket0_R\bra0$ after every loop, telescoping the reduced CPTP maps instead
sums $N$ one-loop bounds, giving $N\eta_H^2/2$. A constant interaction-frame
error $K=k\sigma_x$ has $a(T)=\cos(kT)$ and attains the coefficient
$1/2$ to leading order; without resets the accumulated angle is $NkT$.
In contrast, $K=k\ket0\bra0$ gives
$|a(T)-1|=2|\sin(kT/2)|=O(|kT|)$ and violates the zero-expectation
hypothesis. Thus cyclicity or small error alone cannot justify the
quadratic estimate.

\section{Entropy distance at the Bell crossing}
\label{dyn:bell_entropy_subsection}

For the singlet trajectory in Eq.~\eqref{dyn:werner_solution}, sample
the four input pairs uniformly and fix
$A_0=-\sigma_z$, $A_1=-\sigma_x$,
$B_0=(\sigma_z+\sigma_x)/\sqrt2$, and
$B_1=(\sigma_z-\sigma_x)/\sqrt2$. For outcomes $a,b\in\{-1,1\}$,
\begin{equation}
 P_\zeta(a,b|x,y)=\frac{1+ab(-1)^{xy}\zeta}{4},\qquad
 \zeta=\frac{v}{\sqrt2},\qquad S=4\zeta.
 \label{dyn:symmetric_behavior}
\end{equation}
Let $\mathcal L_{\mathrm{Bell}}$ be the Bell-local polytope in this
binary two-setting scenario. Define the relative-entropy distance
with the stated input distribution by
\begin{equation}
 \mathcal N(P)=\min_{Q\in\mathcal L_{\mathrm{Bell}}}
 \frac14\sum_{x,y,a,b}P(a,b|x,y)
 \ln\frac{P(a,b|x,y)}{Q(a,b|x,y)}.
 \label{dyn:behavior_distance}
\end{equation}
This statistical strength~\cite{VanDamGillGrunwald2005} fixes the
measurements and input frequencies, without protocol or multi-copy
optimization. With $d(p\Vert q)$ denoting binary relative entropy in nats,
\begin{equation}
 \mathcal N(P_\zeta)=
 \begin{cases}
 d((S+4)/8\Vert3/4),&S>2,\\
 0,&S\le2.
 \end{cases}
 \label{dyn:exact_behavior_distance}
\end{equation}
Coarse-graining records into the CHSH win event $ab=(-1)^{xy}$ and
its complement gives win probability $(S+4)/8$, versus at most $3/4$
locally; data processing yields the lower bound above the facet.
The uniform mixture of its eight saturating deterministic strategies,
$P_{1/2}$, attains it because the conditional record distributions given
win or loss match those of $P_\zeta$.
For $0\le\zeta\le1/2$, interpolation between $P_0$ and $P_{1/2}$ proves
locality.

Consequently, from the violating side,
$\mathcal N=(S-2)^2/24+O((S-2)^3)$.
If $\Gamma$ is twice continuously differentiable near a crossing
$t_B=t_{\mathrm{CHSH}}$ with $R_B=\dot\Gamma(t_B)>0$, then
\begin{equation}
 \mathcal N(P_t)=\frac{R_B^2}{6}(t_B-t)^2
 +O((t_B-t)^3),\qquad t\uparrow t_B.
 \label{dyn:entropy_time_onset}
\end{equation}
Indeed $S(t)-2=2R_B(t_B-t)+O((t_B-t)^2)$.
The Hessian $g_{\zeta\zeta}=1/(1-\zeta^2)$ equals $4/3$ at the facet
and fixes the quadratic coefficient, even though this one-dimensional
family has zero intrinsic Riemann curvature.

\section{A sufficient lifetime bound from entropy mixing}
\label{dyn:mixing_subsection}

A second route applies beyond the isotropic model. Consider a
finite-dimensional CPTP semigroup with generator $\mathcal G$ and
faithful stationary state $\tau$. For faithful $\rho$ define
\begin{equation}
 \Sigma(\rho)=-\Tr[\mathcal G(\rho)(\ln\rho-\ln\tau)].
 \label{dyn:entropy_production}
\end{equation}
Stationarity and data processing \cite{Lindblad1975} give $\Sigma\ge0$ and
$\dot{\DU}(\rho_t\Vert\tau)=-\Sigma(\rho_t)$.
A further inequality
$\Sigma(\rho)\ge2\alpha\DU(\rho\Vert\tau)$, $\alpha>0$, gives
\begin{equation}
 \DU(\rho_t\Vert\tau)\le D_0e^{-2\alpha t},\qquad
 D_0=\DU(\rho_0\Vert\tau).
 \label{dyn:entropy_decay}
\end{equation}
The contraction rate is an additional generator hypothesis
\cite{Hanson2020,GauvinCompletion2026}. Faithful approximation extends
the estimate to singular initial states, with finite $D_0$ because
$\tau$ is faithful.

\begin{proposition}[Finite separability certificate]
\label{dyn:mixing_proposition}
Assume Eq.~\eqref{dyn:entropy_decay} with a product stationary state
$\tau=\tau_A\otimes\tau_B$ and
$m_j=\lambda_{\min}(\tau_j)>0$. For $D_0>0$, every time satisfying
\begin{equation}
 t\ge\frac1\alpha
 \left[\ln\frac{\sqrt{2D_0}}{m_A m_B}\right]_+
 \label{dyn:separability_time}
\end{equation}
gives a separable $\rho_t$ across $A:B$. If $D_0=0$, the initial state
is already the stationary product state.
\end{proposition}
\begin{proof}
Quantum Pinsker and Eq.~\eqref{dyn:entropy_decay} imply
\begin{equation}
 \|\rho_t-\tau\|_2\le\|\rho_t-\tau\|_1
 \le\sqrt{2D_0}e^{-\alpha t}.
 \label{dyn:pinsker_bound}
\end{equation}
Here $\|\cdot\|_2$ is the Hilbert-Schmidt norm.
Lemma 2.2 of Ref.~\cite{Hanson2020} makes the ball of radius $m_A m_B$
about $\tau_A\otimes\tau_B$ separable, proving
Eq.~\eqref{dyn:separability_time}.
\end{proof}

The certificate is conservative; a rank-deficient limiting state,
protected sector, or entangled Gibbs reference can fail its hypotheses.

An explicit rate is available for the replacement model, $\kappa_R>0$:
\begin{align}
 \mathcal G_R(\rho)&=\kappa_R(\tau\Tr\rho-\rho),\nonumber\\
 \rho_t&=e^{-\kappa_Rt}\rho_0+(1-e^{-\kappa_Rt})\tau.
 \label{dyn:reset-model}
\end{align}
This semigroup resets the joint preparation to the product $\tau$ at
Poisson-distributed times. For faithful $\rho$,
$\Sigma=\kappa_R[\DU(\rho\Vert\tau)+\DU(\tau\Vert\rho)]
\ge\kappa_R\DU(\rho\Vert\tau)$, so $\alpha=\kappa_R/2$ is valid.
Joint convexity also gives the integrated bound for singular inputs.
For two qubits with $\tau=\id_4/4$ and a Bell input, $D_0=\ln4$, giving
\begin{equation}
 t_{\mathrm{cert}}=\frac{2}{\kappa_R}\ln\bigl(4\sqrt{2\ln4}\bigr)
 \simeq\frac{3.792}{\kappa_R},
 \label{dyn:reset-certificate}
\end{equation}
whereas the exact trajectory becomes separable at $(\ln3)/\kappa_R$. These simultaneous product resets use a shared classical clock, unlike the independent thermal-relaxation generator of Sec.~\ref{th:sec:thermal}.

For comparison, entropy contraction toward an arbitrary separable
reference only gives
\begin{equation}
 S_{\max}(\rho_t)\le
 \min\{2\sqrt2,\,2+4\sqrt{D_0}e^{-\alpha t}\},
 \label{dyn:asymptotic_chsh_bound}
\end{equation}
by the CHSH operator norm and Pinsker. This bound stays above $2$ at
finite time. The finite certificate instead uses the interior separable
ball around a faithful product state and concerns the specified $A:B$
trajectory, not entanglement breaking against every reference system.

\clearpage
\bibliography{references}
\end{document}